\documentclass[unsortedaddress,twocolumn,eqsecnum,prb,floatfix]{revtex4}  

\usepackage[utf8]{inputenc}
\usepackage[dvips]{graphicx}
\usepackage{afterpage}
\usepackage{color}
\usepackage{csquotes}
\usepackage{amsmath}
\usepackage{cancel}
\usepackage{float}
\usepackage{hyperref}

\begin{document}
\title{Local and Charge-Transfer Excitation of Pentacene-Buckminsterfullerene complexes}
\author{Ala Aldin M.Hani M. Darghouth}
\author{Omar A.Shareef}
\author{ Firas A. AL-Lolage}
\affiliation{Department of Chemistry,\\
           College of Science, \\
           University of Mosul}
\author{Mark E.\ Casida}
\affiliation{
        D\'epartement de Chimie Mol\'eculaire (DCM, UMR CNRS/UJF 5250),\\
        Institut de Chimie Mol\'eculaire de Grenoble (ICMG, FR2607),\\
        Universit\'e Grenoble-Alpes),\\
        301 rue de la Chimie, BP 53,\\
        F-38041 Grenoble Cedex 9, France}

\begin{abstract}

The charge transfer state, the local excited state on pentacene, and the local excited state on buckminsterfullerene have been studied for four models of the pentacene-buckminsterfullerene organic solar cell. These models have different interface configurations between the donor and the acceptor. The study has been done by using different functionals and time-dependent density functional theory-namely, the long range-separated hybrid with coulomb-attenuating method approach (CAM-B3LYP functional), the popular B3LYP (Becke, three-parameter, Lee-Yang-Parr) exchange-correlation functional, and the density functional tight binding (DFTB) method. The charge transfer state energy obtained by using CAM-B3LYP without optimally tuned range-separated hybrid parameters are very close to those obtained by using a many-body dispersion-corrected, optimally tuned range-separated hybrid functional (OPT-$\omega$B97XD). Both the B3LYP functional and the DFTB method fail to describe correctly the charge transfer energy for all models. Each of them is underestimating  around 1 eV compared with range-separated hybrid functional.

\end{abstract}
\maketitle
\tableofcontents
\section{Introduction}
Organic solar cells have been extensively studied in recent years because of their potential for being a low-cost photovoltaic technologies.  Nowadays, organic solar cell efficiency as of late has come to up to 20\% \cite{JJCYYZY23, KJRZ25, FYZZ26}. It will standout as one of the most encouraging renewable energy sources from the perspective of, low production costs in high volumes, flexibility, molecular engineering, high optical absorption coefficient, light weight, large area applications, and printability \cite{BFY23, BSF09, FCCHJOTS07}. Keeping in mind that the end goal is to encourage progress in efficiency, significant effort must be made to improve our understanding of factors governing organic solar cell performance to make it economically more fruitful.
Short-circuit current density (Jsc) and open-circuit voltage (Voc) are two of the essential parameters for efficiency \cite{TOSPHF06}. These parameters can give high power conversion efficiency of the organic solar cell through high open-circuit voltages, short-circuit currents, and large fill-factors (FFs)\cite{PRB+09,HTYHYL06}. Numerous experimental efforts have been undertaken to clarify the relationship between open-circuit voltage and the difference between the highest
occupied molecular orbital (HOMO) of the donor and lowest unoccupied molecular orbital (LUMO) of the
acceptor \cite{MEMNKGX24, SMKDWHB06, BCMSFRSH01}.
Upon deeper consideration, the picture is not so simple. The function and efficiency of the organic solar cell are basically dependent on the electronic structure of the interface between the donor and the acceptor in light of the fact that exciton separation into electrons and holes happens here. So charge separation at organic donor/acceptor heterojunctions is a key element in the fruitful design of organic solar cells \cite{CD10, BNCC09, ZYM09}.
Understanding the mechanism of the charge separation in the organic solar cell still presents a significant challenge. There are a few speculations for the mechanism. Some experiments seem to indicate that the mechanism happens through charge transfer (CT) complex states (exciplex) and this is most clear in the materials where the exciplex is lower in energy  than the charge-separated state \cite{MDKSGMMCBF+03}. Likewise the specific mechanism of this separation presumably includes a moderately abnormal state of delocalization of hot CT states \cite{BRPVPNCBF12}.
The charge transfer state is portrayed as an excited state where the elevated electron is moved to the closest neighboring molecular site, yet at the same time remains coulombically associated with its parent hole \cite{PS82}.

So, a key part for the capacity of organic solar cells is played by the charge transfer state. 
Diverse terms are utilized to allude to this intermediate state between the exciton and the completely isolated charge. A few illustrations are: geminate pair, bound polaron pair, and exciplex. In any case, these diverse terms allude to distinctive  physical properties. The \enquote{charge transfer exciton} is attainable from the ground state,\cite{PS82} while the terms exciplex, geminate pair, and polaron pair are generously synonyms \cite{S97} In this content we will allude to a \enquote{charge transfer state}.  
In this manner, getting a precise depiction of these states is basic to achieve a complete comprehension of the working mechanism of the organic solar cells. A noteworthy snag in the demonstrating of CT states emerges from the powerlessness of traditional density functional theory (DFT) methods to portray their nature precisely. Basically, these troubles can be followed back to two noteworthy and interrelated inadequacies of traditional functionals, in particular one of them, the poor depiction of the $1/r$ ($r$ = length) reliance for the energy of the CT states at large intermolecular partitions because of the local or semilocal nature of the functionals; and second, the poor depiction of the basic gap in molecular systems because of the electron self-interaction error \cite{ZSYACB14}.
Hybrid functionals are advantageous for molecular density functional theory (DFT)applications, since they often give results near the experimental methods \cite{KH00}. They incorporate ordinarily around 20\% of Hartree-Fock (HF) exchange in addition to the traditional density functional exchange \cite{ADB93}. The achievement of hybrid functionals can be ascribed to a halfway diminishment of the self-interaction error, without significantly diminishing the balance between exchange and correlation.
Sadly, a uniform blending of a little divide of Hartree-Fock exchange is inadequate to right the majority of the unfavorable components of the DFT functionals, similar to the need of derivative discontinuity or the wrong asymptotic conduct of the exchange potential. Rather than the right $1/r$ asymptotic type of the  hybrid exchange potential potential one acquires $a/r$, where a is the extent of the Hartree-Fock exchange in the hybrid. Asymptotically corrected  exchange potentials may enhance orbital energies and the state of orbitals, however they do not give access to total energies, since the relating practical remains obscure \cite{ICGJGA05}.
An alternate methodology consists of restricting the long range correction of the exchange energy density and of the potential, by utilizing a suitable function to divide the Coulomb operator into short and long range parts. This method, called range-separated or long-range corrected xc functionals was utilized by Savin and his collaborators \cite{TLHSHWAS97, AS96} to allay the aforementioned troubles. In particular, the thought is to divide the electron-electron interaction into short-range (sr) and long-range (lr) contributions

\begin{equation}
\frac{1}{r_{12}}= \frac{1- \text{erf}(\omega r_{12})}{r_{12}} + \frac{\text{erf}(\omega r_{12})}{r_{12}}
\label{eq:1}
\end{equation}
where the long range part is dealt with precisely, while the short-range part offers ascend to an altered pure density functional. In Eq. (\ref{eq:1}), the expression \enquote{erf} gives the error function and $\omega$ is an empirical parameter. The subsequent hypothesis may be seen as a special type of hybrid functionals with settled weighting coefficients of density functional and HF exchange. Also, the error cancellation of density functionals for exchange and correlation is kept in the short range, while the self-interaction error is removed asymptotically through the long-range contribution.

In this work, a hybrid exchange-correlation named CAM-B3LYP is utilized. It consolidates the hybrid characteristics of B3LYP and long-range correction displayed by Tawada \textit{et al}. \cite{TYTTYSYTHK04}. The CAM-B3LYP functional involves 0.19 Hartree-Fock  in addition to 0.81 Becke 1988 (B88) exchange interaction at short-rang, and 0.65 HF in addition to 0.35 B88 at long-range. The range separation uses the error function with parameter 0.33 \cite{TYDPTNCH04}. 

\begin{figure}[H]
\includegraphics[scale=0.35]{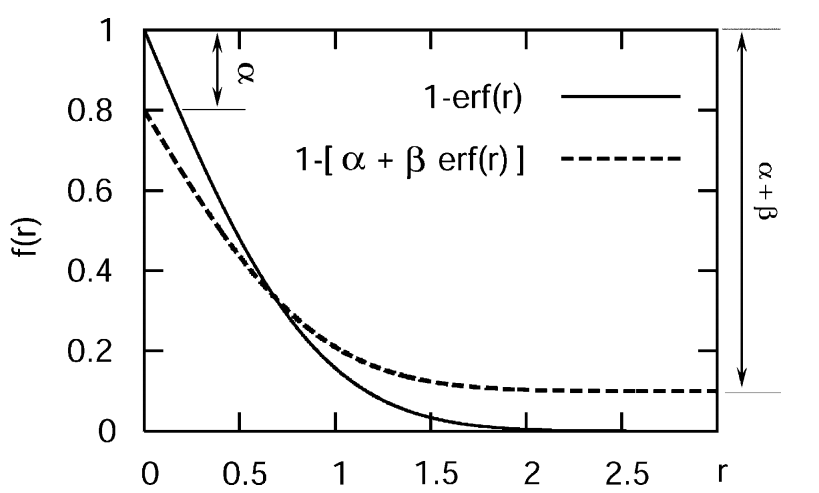}
\caption{\label{fig:1} Plots for $f(r)=1-erf(\omega r)$ and $f(r)=1-[\alpha + \beta erf (\omega r)]$ taken from ref. \cite{TYDPTNCH04} }
\end{figure}

CAM-B3LYP generalize the form of Eq. (\ref{eq:1}) using two extra parameters $\alpha$ and $\beta$ as, 

\begin{equation}
\frac{1}{r_{12}}= \frac{1-[\alpha+\beta \cdot \text{erf}(\omega r_{12})]}{r_{12}} + \frac{\alpha+\beta \cdot \text{erf}(\omega r_{12})}{r_{12}},
\label{eq:2}
\end{equation}
the parameter $\alpha$ allows us to incorporate the HF exchange contribution over the entire rang by a component of $\alpha$, and the parameter $\beta$ allows us to incorporate the DFT counterpart  over the entire range by a component of 1-($\alpha$+$\beta$). Fig. \ref{fig:1} outlines the schematic plots of two functions, Eqs. (\ref{eq:1}) and (\ref{eq:2}).
Yanai and \textit{et al.} noticed that the generally utilized hybrid functional B3LYP \citep{Becke93} takes CAM potential apportioning of Eq. (\ref{eq:1}) with $\alpha$ = 0.2 and $\beta$=0 for the blending of Slater exchange $E_X^{Slater}$  and the HF  exchange $E^{HF}_ X$. i.e.,

\begin{equation}
E^{B3}_X= (1-\alpha)E_X^{Slater}+\alpha E^{HF}_ X +\Delta E^{B88}_ X,
\label{eq:3}
\end{equation}
where the extra term $\Delta E^{B88}_ X$ is Becke's 1988 gradient
correction for exchange \citep{ADB88} with the semiempirical parameter 
$c^{B88}$= 0.72, which Becke got by a linear least-square fit to experimental data for a single atom \cite{Becke93}.
The formalism and the fundamentals of the theoretical methods, B3LYP and DFTB have been discussed several times in the previous reports.

In this report, I will try to deal with two important questions, one of them is:
What is the difference in the behavior between functionals without range-separated hybrid (like B3LYP) and functional with range-separated hybrid (like CAM-B3LYP) in the calculating of the energy of the charge-transfer state, the local excited state on pentacene, and the local excited state on fullerene?
The second question is: How these functionals with and without range-separated hybrid succeed in describing the charge transfer energy between the donor and acceptor, in addition to studying the effect of the interface configuration between the donor and the acceptor on the charge transfer state with these functionals.

\section{Model Systems}
\begin{widetext}
\begin{figure}[H]
\begin{center}
\includegraphics[scale=0.4]{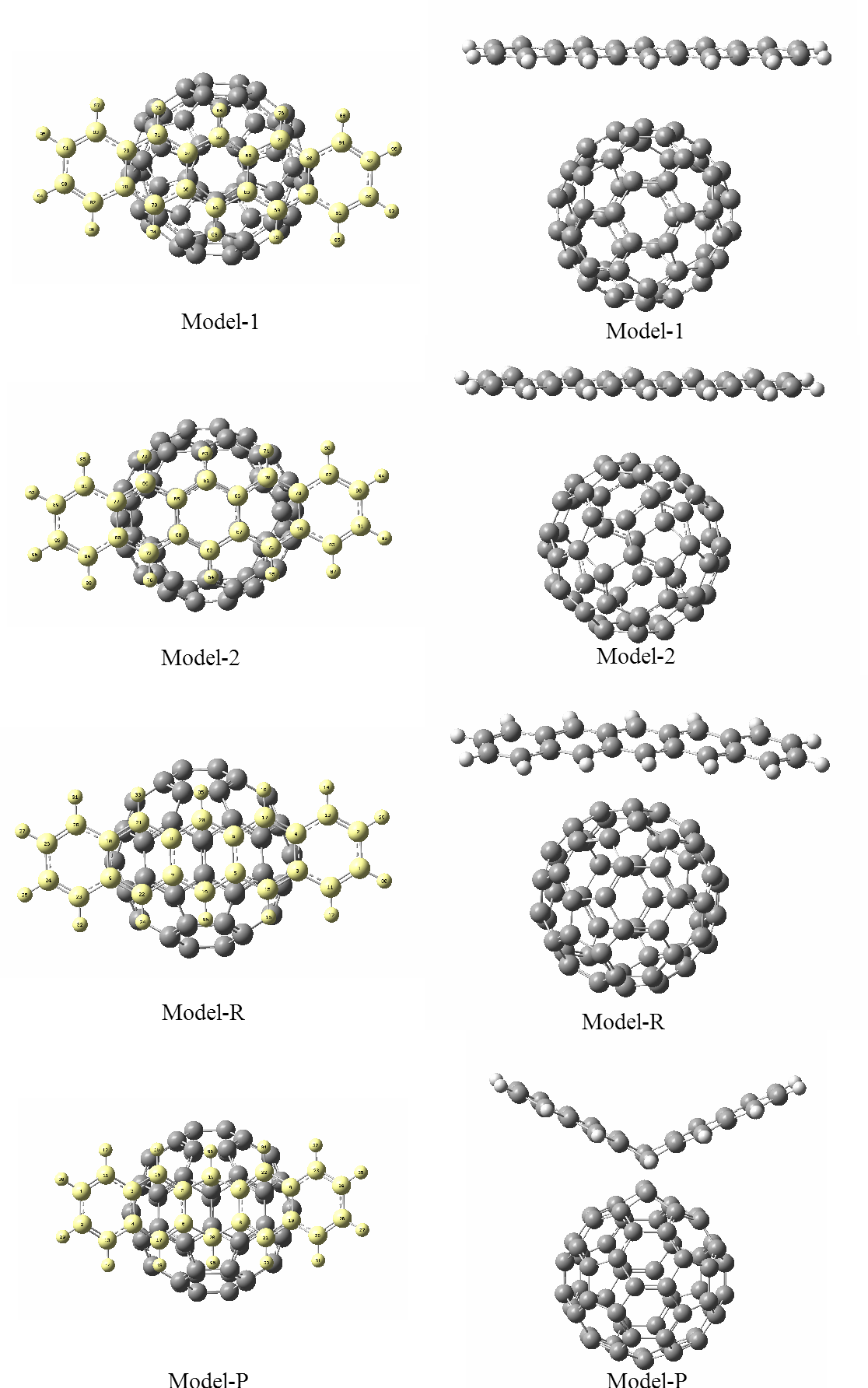}
\caption{\label{fig:2}Models of relevance for pentacene/buckminsterfullerene organic solar cells studied in this work.}
\end{center}
\end{figure}
\end{widetext}

The donor/acceptor interface geometry in organic solar cells plays a significant role in both exciton separation and the charge recombination that take place after exciton separation. The exciton separation which is responsible for creating the photocurrent should be amplified whereas the charge recombination diminishes the photocurrent should be minimized. In addition, the size of the saturation current in the dark which should be minimized to expand the open-circuit voltage \cite{PYK08, PBFT09, YCB09}.
For example the pentacene long axis oriented perpendicular to the acceptor orientation is less good for hole transport as intermolecular electronic couplings are weaker along the axis \cite{CCDOSB07}. As a result, this intermolecular orientation is less good for effective exciton separation \cite{YCB09}.
In the case of pentacene/buckminsterfullerene,  Yuanping Yi \textit{et al.} \cite{YCB09} mentioned, pentacene/buckminsterfullerene pairs can be relied upon to be in a parallel orientation, a design more suitable for strong electronic coupling between the pentacene and $\text{C}_{\text{60}}$ electronic states and, subsequently, for effective exciton separation. 
So, four models of pentacene and buckminsterfullerene have been constructed in a parallel orientation. The first model (mode-1) has the central ring of pentacene centered over a 5-membered ring of $\text{C}_{\text{60}}$. The second model (model-2) has the central ring of pentacene centered over a 6-membered ring of $\text{C}_{\text{60}}$. These two models have been studied by Zhang \textit{et al.} \cite{ZSYACB14} with others functionals. Here, we want to study these models with other functionals as well as, in the near future we want to study the charge transfer state in these models by using range-separated hybrid DFTB method and compare the results with the other functionals. The third (Model-R) and the fourth (Model-P) represent the reactant and the product of the pentacene and buckminsterfullerene reaction respectively where these two models, the central 6,13 carbons in pentacene centred over  a the $\pi$ bond between two fused 6-membered rings in $\text{C}_{\text{60}}$ . Also, the four models show us different $\pi$ orbitals stacking   between the donor and acceptor. All models are shown in fig. \ref{fig:2}.

\onecolumngrid
\twocolumngrid


\section{Computational Details }
Complete geometry optimizations were carried out for each molecular species with B3LYP functional \cite{SDCF94} but initially, the geometries of the molecules had been taken from Crystallography Open Database (COD)\cite{SSMHADBN07, DM94}.

The molecular geometries  that have been optimized by using  B3LYP/6-31G(d,p). These optimized geometries consequently have been used to construct the models (model-1, model-2, model-R, and model-P).

D3 version of Grimme's \cite{GAEK10} dispersion with the original D3 damping function have been used for 
constructing the potential energy surfaces of each of the model setups utilizing CAM-B3LYP \cite{TYDPTNCH04}functional.

Single point density functional (DFT) and time-dependent DFT \cite{CJCS98, BA96}  calculations by using B3LYP and CAM-B3LYP functionals have been carried out by using 6-31G(d,p) basis set and the {\sc Gaussian 09} program \cite{Gaussian09} for the individual pentacene and Buckminsterfullerene molecules as well as for all models.

Time-dependent DFTBA \cite{TSZCFB11} calculations were carried out using the {\sc Gaussian 09} program \cite{Gaussian09} package.  DFTB uses the tabulated matrix elements as in the original implementation of Elstner and coworkers \cite{PFKSK95, EPJEHFSS98}.

\section{Results}
\subsection{Potential energy curve of the models systems}
\begin{widetext}
\begin{figure}[H]
\begin{center}
\includegraphics[scale=0.45]{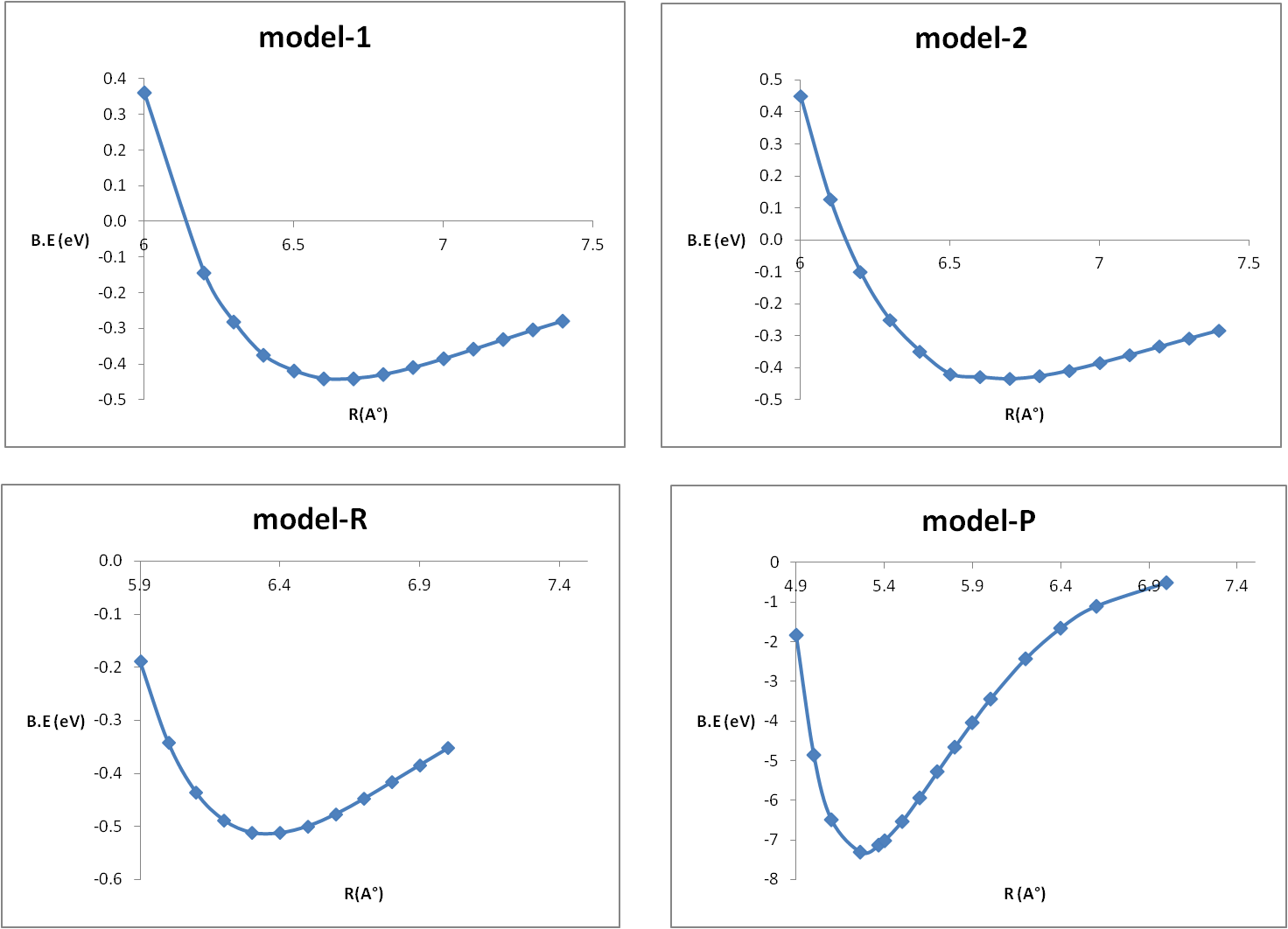}
\caption{\label{fig:3}Potential energy surfaces in eV of the four models by using CAM-B3LYP with empirical dispersion GD3.} 
\end{center}
\end{figure}
\end{widetext}

Due to the importance of the orientation of each molecule in the model and the separated distance ($1/r$ ) between the two molecules in the model on the charge transfer states in addition to the degree of their blending with the local excitations either in pentacene or buckminsterfullerene, the potential energy curves have been studied for the four models as shown in the Fig. \ref{fig:3}. The distance between the two molecules centres has been varied in one-dimension using frozen geometries. In the model-1 (the central ring of pentacene centred over a 5-membered ring of $\text{C}_{\text{60}}$) and model-2 (the central ring of pentacene centered over a 6-membered ring of $\text{C}_{\text{60}}$), the distance between the two molecules' centres in the model has been varied in one-dimensional from (6 - 7.4 \AA ) whereas the model-R from (5.9 - 7 \AA ) and model-P from (4.9 - 7.0 \AA ). The potential energy curve of the model-1 using CAM-B3LYP shows the minimum of the ground state energy at 6.6 \AA \ with binding energy around 0.4411 eV using D3 version of Grimme's \cite{GAEK10} dispersion with the original D3 damping function, whereas the potential energy curve of the model-2 by using the same functional shows the minimum of the ground state energy at 6.7 \AA \ with binding energy around 0.4346 eV by using the same empirical dispersion GD3.
But this binding energy reduced approximately by a third when we use counterpoise corrections \cite{BB07, SDD96} (basis set superposition error) to reach 0.3041 eV in model-1 and 0.3045 eV in model-2.
As you see, there is a slight difference in the binding energy between the two models. Also, if we compare our results with those obtained from Zhang \textit{et al.} \cite{ZSYACB14}, the difference in the binding energy around 0.2 eV this difference can be attributed to, the range separated correction which has been used in our functional but not by Zhang \textit{et al.} also, the counterpoise corrections that missed by Zhang \textit{et al.} calculations.

\onecolumngrid
\twocolumngrid

In the model-R, the potential energy curve obtained by using CAM-B3LYB with empirical dispersion (GD3) shows the minimum ground state energy at 6.4 \AA \ with binding energy around 0.5122 eV. Whereas, the potential energy curve of the model-P with the same functional shows the minimum ground state energy at 5.27 \AA \ with binding energy 7.31 eV fourteen times more than model-R.
But when we use counterpoise corrections to cancel the basis set superposition error the binding energy will be 0.3611 eV and 6.996 eV for model-R and model-P respectively.
The binding energy of model-P is twenty-three times greater than the binding energy of both model-1 and model-2 and fourteen times greater than the binding energy of model-R this is normal given to the nature of $\pi$ bond  between pentacene and $\text{C}_{\text{60}}$ in the model-P compared with the van der Waals forces that bind between the molecules in the model-1, model-2, and model-R.
The CAM-B3LYP gives $\pi$-stacking separations between the focal ring of pentacene and the closest ring of $\text{C}_{\text{60}}$ in model-1 and model-2 somewhere around 3.28 and 3.44 \AA \ respectively. Whereas, the  $\pi$-stacking distance between the central 6,13 carbons in pentacene and  the $\pi$ bond between two fused 6-membered rings in $\text{C}_{\text{60}}$ in model-R around 3.11 \AA \ and the length of $\pi$ bonds between the central 6,13 carbons in pentacene and the two carbons of fused  6-membered rings in $\text{C}_{\text{60}}$ in model-P is 1.495 \AA.

\onecolumngrid
\begin{widetext}
\begin{table}[htbp]
  \centering
  \caption{The counterpoise corrected  energy, the basis set superposition error energy, and binding energy after correction for all models.}
    \begin{tabular}{rrrrr}
    \toprule
    \multicolumn{1}{c}{\textbf{R}} & \multicolumn{1}{c}{\textbf{E.W.C.C}} & \multicolumn{1}{c}{\textbf{C.C.E}} & \multicolumn{1}{c}{\textbf{ BSSE energy}} & \multicolumn{1}{c}{\textbf{B.E (e.V)}} \\    
    \toprule
    \multicolumn{5}{c}{\textbf{model-1}} \\    
    \multicolumn{1}{c}{6.4} & \multicolumn{1}{c}{-3131.3701} & \multicolumn{1}{c}{-3131.3640} & \multicolumn{1}{c}{0.0061} & \multicolumn{1}{c}{0.210} \\
    \multicolumn{1}{c}{6.5} & \multicolumn{1}{c}{-3131.3717} & \multicolumn{1}{c}{-3131.3663} & \multicolumn{1}{c}{0.0056} & \multicolumn{1}{c}{0.271} \\
    \multicolumn{1}{c}{6.6} & \multicolumn{1}{c}{-3131.3725} & \multicolumn{1}{c}{-3131.3675} & \multicolumn{1}{c}{0.0050} & \multicolumn{1}{c}{0.304} \\
    \multicolumn{1}{c}{6.7} & \multicolumn{1}{c}{-3131.3725} & \multicolumn{1}{c}{-3131.3679} & \multicolumn{1}{c}{0.0046} & \multicolumn{1}{c}{0.317} \\
    \multicolumn{5}{c}{\textbf{model-2}} \\
    \multicolumn{1}{c}{6.4} & \multicolumn{1}{c}{-3131.3691} & \multicolumn{1}{c}{-3131.3628} & \multicolumn{1}{c}{0.0063} & \multicolumn{1}{c}{0.178} \\
    \multicolumn{1}{c}{6.5} & \multicolumn{1}{c}{-3131.3717} & \multicolumn{1}{c}{-3131.3662} & \multicolumn{1}{c}{0.0055} & \multicolumn{1}{c}{0.270} \\
    \multicolumn{1}{c}{6.6} & \multicolumn{1}{c}{-3131.3721} & \multicolumn{1}{c}{-3131.3668} & \multicolumn{1}{c}{0.0053} & \multicolumn{1}{c}{0.286} \\
    \multicolumn{1}{c}{6.7} & \multicolumn{1}{c}{-3131.3723} & \multicolumn{1}{c}{-3131.3675} & \multicolumn{1}{c}{0.0048} & \multicolumn{1}{c}{0.305} \\
    \multicolumn{1}{c}{6.8} & \multicolumn{1}{c}{-3131.3720} & \multicolumn{1}{c}{-3131.3677} & \multicolumn{1}{c}{0.0043} & \multicolumn{1}{c}{0.309} \\
    \multicolumn{5}{c}{\textbf{model-R}} \\
    \multicolumn{1}{c}{6.2} & \multicolumn{1}{c}{-3131.3784} & \multicolumn{1}{c}{-3131.3717} & \multicolumn{1}{c}{0.0067} & \multicolumn{1}{c}{0.305} \\
    \multicolumn{1}{c}{6.3} & \multicolumn{1}{c}{-3131.3793} & \multicolumn{1}{c}{-3131.3732} & \multicolumn{1}{c}{0.0061} & \multicolumn{1}{c}{0.347} \\
    \multicolumn{1}{c}{6.4} & \multicolumn{1}{c}{-3131.3793} & \multicolumn{1}{c}{-3131.3738} & \multicolumn{1}{c}{0.0056} & \multicolumn{1}{c}{0.361} \\
    \multicolumn{1}{c}{6.5} & \multicolumn{1}{c}{-3131.3788} & \multicolumn{1}{c}{-3131.3738} & \multicolumn{1}{c}{0.0050} & \multicolumn{1}{c}{0.363} \\
    \multicolumn{5}{c}{\textbf{model-P}} \\
    \multicolumn{1}{c}{5.1} & \multicolumn{1}{c}{-3131.3889} & \multicolumn{1}{c}{-3131.3759} & \multicolumn{1}{c}{0.0131} & \multicolumn{1}{c}{6.136} \\
    \multicolumn{1}{c}{5.27} & \multicolumn{1}{c}{-3131.4188} & \multicolumn{1}{c}{-3131.4075} & \multicolumn{1}{c}{0.0113} & \multicolumn{1}{c}{6.997} \\
    \multicolumn{1}{c}{5.37} & \multicolumn{1}{c}{-3131.4129} & \multicolumn{1}{c}{-3131.4025} & \multicolumn{1}{c}{0.0104} & \multicolumn{1}{c}{6.861} \\
    \multicolumn{1}{c}{5.4} & \multicolumn{1}{c}{-3131.4084} & \multicolumn{1}{c}{-3131.3983} & \multicolumn{1}{c}{0.0101} & \multicolumn{1}{c}{6.747} \\
    \toprule
    \multicolumn{5}{l}{R: The distance between the two molecules center in Angstroms} \\
    \multicolumn{5}{l}{E.W.C.C : The total energy in eV with out counterpoise corrected energy } \\
    \multicolumn{5}{l}{W.C.C.E : Counterpoise corrected energy } \\
    \multicolumn{5}{l}{ BSSE:  Basis set superposition error energy} \\
    \multicolumn{5}{l}{ B.E:  Binding energy in eV} \\
    \toprule
    \end{tabular}%
  \label{tab:1}%
\end{table}%
\end{widetext}
\twocolumngrid 

\subsection{Model-1, the central ring of pentacene centered over a 5-membered ring of $\text{C}_{\text{60}}$}
\begin{figure}[H]
\begin{center}
\includegraphics[width=\linewidth]{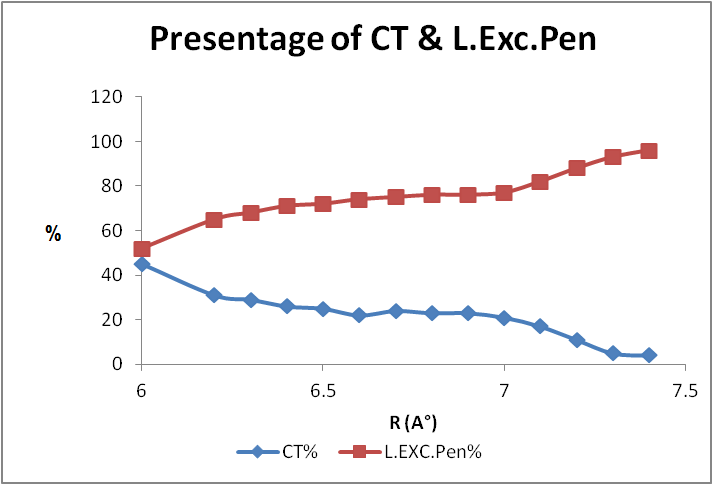}
\caption{\label{fig:14}The relationship between the distance and the percentage of each state at the same excitation by using CAM-B3LYP functional in model-1.} 
\end{center}
\end{figure}

Calculation of the excited states by using range separated hybrid have been done with CAM-B3LYP functional with 6-31G(d,p) basis set using D3 version of Grimme \cite{GAEK10} dispersion with the original D3 damping function. The calculations of the excited states in model-1 revealed that the lowest excited state at 2.247 eV is mostly local excited state on pentacene from HOMO to LUMO+3 with percentage 74\% and mixed with small percentage 22\% of charge transfer state from pentacene to $\text{C}_{\text{60}}$ (from HOMO to LUMO). The percentage of charge transfer state will decrease when increasing the distance between the donor and the acceptor that is a clear evidence that the charge transfer state depends on the strong of $\pi$-orbital stacking between the two fragments. Figure \ref{fig:14} show us the relationship between the distance and the percentage of each state at the same excitation.
The second excited state is mostly charge transfer state with percentage 94\% from HOMO to LUMO+1 and LUMO+5 in addition to a very small percentage of local excitation on fullerene 3\% from HOMO-2 to LUMO +1. The local excited state on fullerene can be seen clearly in the fifth excited state which include transitions from HOMO-1, HOMO-3, HOMO-5, HOMO-4, HOMO-3, and HOMO-2 to  LUMO +2, LUMO, LUMO, LUMO +1, LUMO +2, and LUMO +1 respectively.

In the model-1 also, the calculation of the excited states has been done with B3LYP functional using the 6-31G(d,p) basis set and empirical dispersion. The calculations of the excited states by B3LYP in model-1 revealed that the lowest excited state at 1.269 eV is a pure charge transfer state from pentacene to $\text{C}_{\text{60}}$ ( from HOMO to LUMO+1) with percentage 99\% whereas, the fourth excited state at 1.916 eV is a pure local excited state on pentacene with percentage 100\% from HOMO to LUMO+4 and the local excited state on fullerene can be seen clearly in the fifth excited state  at 2.074 eV which include two transitions from HOMO-1 and HOMO-5  to LUMO and LUMO+5  with presentage 88\% and 8\% respectively.
By using TD-DFTB methods to calculate the excited states, the calculations for model-1 show us  that the lowest excited state at 1.211 eV is a pure charge transfer state from pentacene to $\text{C}_{\text{60}}$ (from HOMO to LUMO) with percentage 100\% whereas, the fourth excited state at 1.833 eV is a pure local excited state on pentacene with percentage 98\% from HOMO to LUMO+3 and the local excited state on fullerene can be seen clearly in the fifth excited state  at 1.902 eV which includes two transitions from HOMO-1 and HOMO-5  to LUMO and LUMO+2  with presentage 74\% and 25\% respectively.
\onecolumngrid
\begin{widetext}
\begin{figure}[H]
\begin{center}
\includegraphics[width=\linewidth]{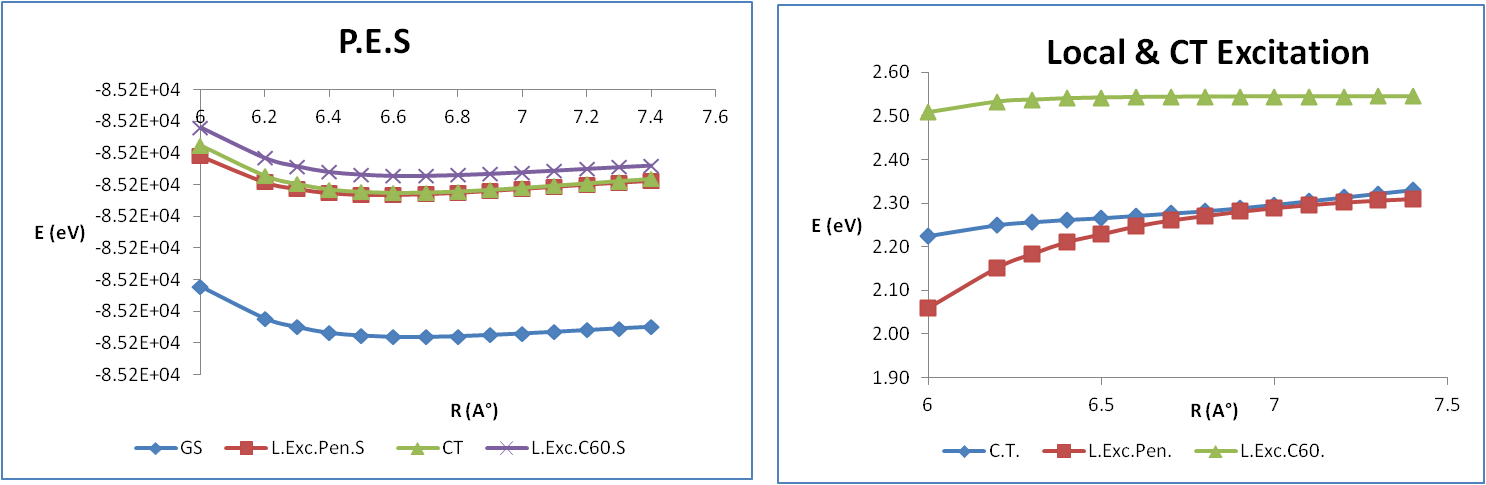}
 \caption{\label{fig:4}Potential energy surfaces in eV of the ground state (GS), local excitation state on pentacene (L. Exc. Pen. S), local excitation state on $\text{C}_{\text{60}}$ (L. Exc. $\text{C}_{\text{60}}$. S), and charge transfer state from pentacene to $\text{C}_{\text{60}}$ in model-1 by using CAM-B3LYP with empirical dispersion GD3.} 
\end{center}
\end{figure}
\end{widetext}
\twocolumngrid

\begin{table}[htbp]
  \centering
  \caption{Local excitation state on pentacene (L. Exc. Pen.), local excitation state on $\text{C}_{\text{60}}$ (L. Exc. $\text{C}_{\text{60}}$. S), and charge transfer state from pentacene to $\text{C}_{\text{60}}$ in model-1 by using different methods. All energies are given in eV.}
    \begin{tabular}{cccc}
 \toprule    
    \textbf{} & \textbf{MODEL\_1} & \textbf{} & \textbf{} \\
    \textbf{Method} & \textbf{L. Exc. Pen.} & \textbf{L. Exc. $\text{C}_{\text{60}}$} & \textbf{C.T.} \\
     \hline
    \textbf{OPT- wB97XD \footnotemark[1]} & \text{2.240} & \text{2.530} & \text{2.310} \\
    \textbf{CAM-B3LYP\_D} & \text{2.247} & \text{2.543} & \text{2.271} \\
    \textbf{B3LYP} & \text{1.916} & \text{2.074} & \text{1.269} \\
    \textbf{TDDFTB} & \text{1.833} & \text{1.902} & \text{1.211} \\
    \toprule
    \end{tabular}%
    \footnotetext[1]{Taken from Ref.~\cite{ZSYACB14}.}
  \label{tab:2}%
\end{table}%

The results in Table \ref{tab:2} show us that the charge transfer state, local excitation state on pentacene, and local excitation state on $\text{C}_{\text{60}}$ calculated in this work by CAM-B3LYP is very close to those calculated by Zhang \textit{et al.} \cite{ZSYACB14} using many-body dispersion-corrected, optimally tuned range-separated hybrid functional (OPT-wB97XD \cite{PAATLK13}) for the same model despite the fact that , we do not use optimized tuned range-separation parameters as they did. But, when we go from range separated hybrid functional to B3LYP functional the difference in the local excited state energy is around 0.4 eV (blueshift) and is around 0.5 eV in the DFTB whereas, the difference is quite enough big in the charge transfer state energy when we go from range separated hybrid functional to the B3LYP functional and DFTB to reach more than 1 eV.

\begin{figure}[H]
\begin{center}
\includegraphics[width=\linewidth]{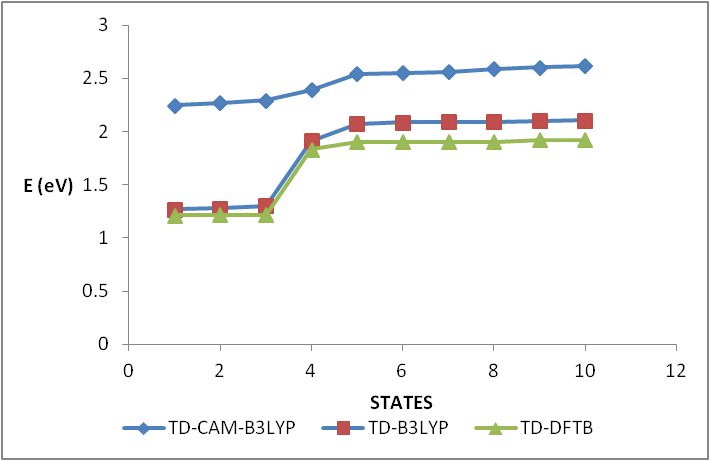}
\caption{\label{fig:5}The difference in energy between different functionals in the local excited states and charge transfer states in model-1.} 
\end{center}
\end{figure}

Fig. \ref{fig:5} show us the difference in energy between different functionals in the local excited states and charge transfer states. Generally, the difference between the B3LYP functional and DFTB for calculating the local excited state energies and the charge transfer state energies  is very small. It is between (0.06- 0.14 eV) when we go from lowest excited state to the highest excited state. But the difference in the energy between the range-separated hybrid functional (CAM-B3LYP) and the other functional without range-separated hybrid for calculating the local excited state energies and charge transfer state energies is about  0.64 eV with B3LYP and 0.77 eV with DFTB.

In the model-1, it is worth mentioning, there is avoided crossing region at around 7 \AA between the potential energy surface of local excited state on pentacene and  the potential energy surface of the charge transfer state as you see in the Fig. \ref{fig:4}.
\onecolumngrid
\begin{widetext}  
\begin{table}[htbp]
  \centering
  \caption{Molecular orbital composition of each fragment in the model-1 by using different functionals.}
    \begin{tabular}{ccccccccccccr}
    \toprule
    \multicolumn{13}{c}{\textbf{MODEL\_1}} \\
     \toprule
    \textbf{} & \multicolumn{12}{c}{\textbf{CAM-B3LYP}} \\
    \text{orbitals} & \text{H-5} & \text{H-4} & \text{H-3} & \text{H-2} & \text{H-1} & \text{H} & \text{L} & \text{L+1} & \text{L+2} & \text{L+3} & \text{L+4} & \multicolumn{1}{c}{\text{L+5}} \\
    \text{pentacene} & 0     & 1     & 5     & 1     & 3     & 98    & 2     & 0     & 0     & 96    & 0     & \multicolumn{1}{c}{1} \\
    \text{fullerene} & 100   & 99    & 95    & 99    & 97    & 2     & 98    & 100   & 100   & 4     & 100   & \multicolumn{1}{c}{99} \\
    \textbf{} & \multicolumn{12}{c}{\textbf{B3LYP}} \\
    \text{pentacene} & 0     & 1     & 2     & 8     & 3     & 97    & 1     & 0     & 1     & 96    & 0     & \multicolumn{1}{c}{1} \\
    \text{fullerene} & 100   & 99    & 98    & 92    & 97    & 3     & 99    & 100   & 99    & 4     & 100   & \multicolumn{1}{c}{99} \\
    \textbf{} & \multicolumn{12}{c}{\textbf{DFTB}} \\
    \text{pentacene} & 0     & 0     & 0     & 0     & 1     & 100   & 0     & 0     & 0     & 100   & 0     & \multicolumn{1}{c}{0} \\
    \text{fullerene} & 100   & 100   & 100   & 100   & 99    & 0     & 100   & 100   & 100   & 0     & 100   & \multicolumn{1}{c}{100} \\
     \toprule
    \end{tabular}%
  \label{tab:3}%
\end{table}%
\end{widetext}
\twocolumngrid

\begin{widetext}
\begin{figure}[H] \begin{center}
    \begin{minipage}[b]{0.45\textwidth}
	\begin{figure}[H] \begin{center}
	\includegraphics[width=\textwidth]{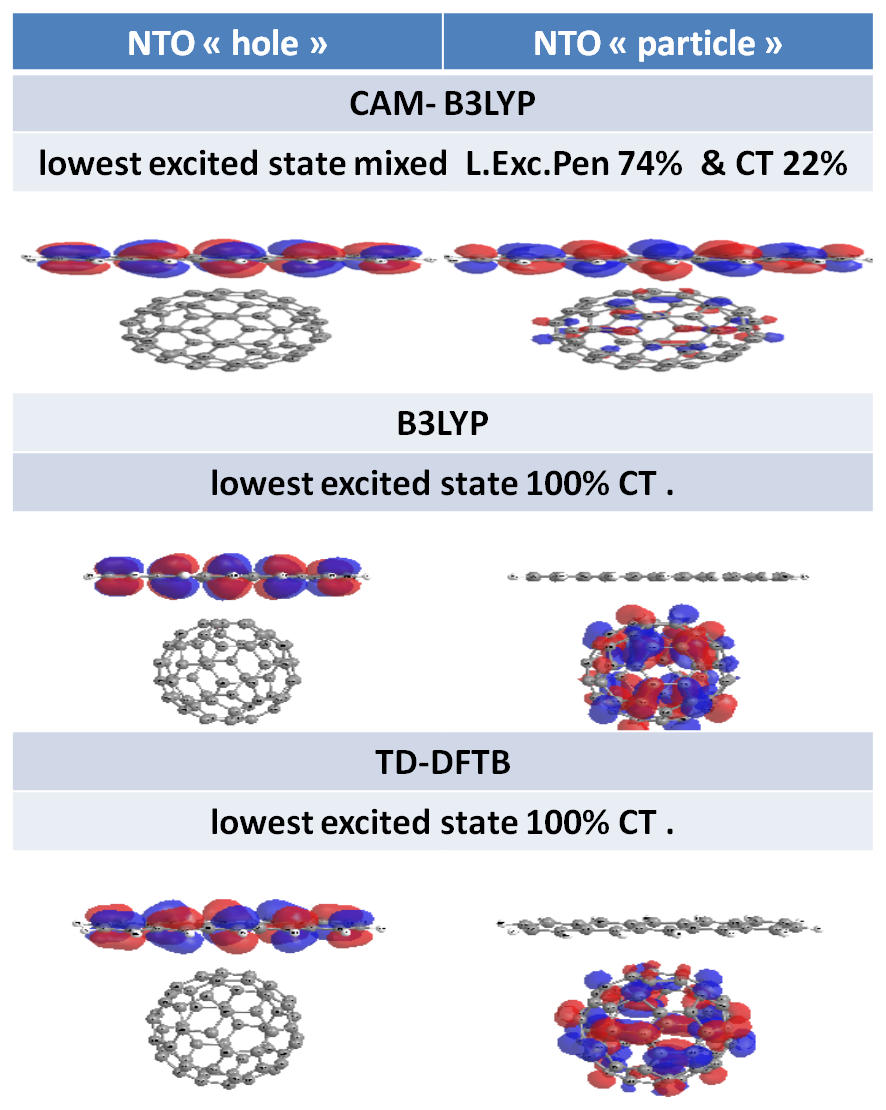}\\
	\scriptsize{a) Charge transfer state}
	\end{center} \end{figure}
    \end{minipage}
    \begin{minipage}[b]{0.45\textwidth}
	\begin{figure}[H] \begin{center}
	\includegraphics[width=\textwidth]{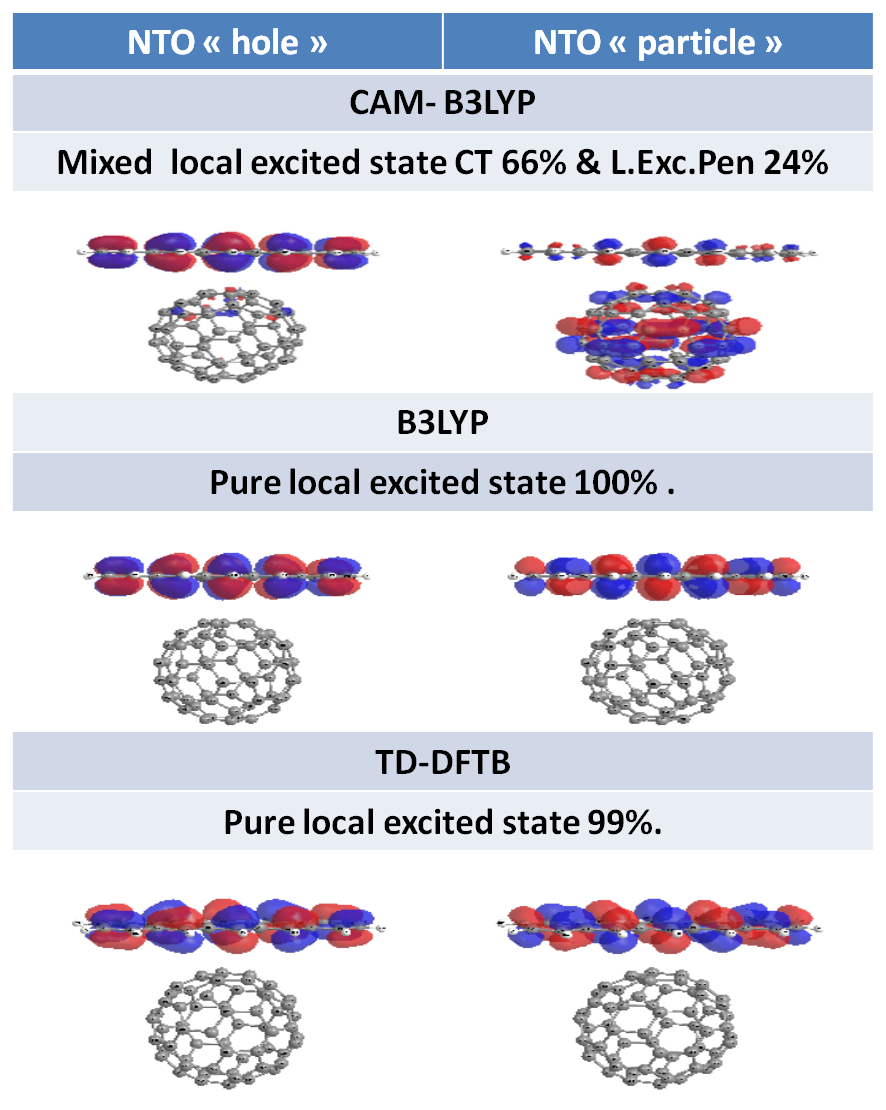}\\
	\scriptsize{b) Local excited state on pentacene}
	\end{center} \end{figure}
     \end{minipage}
     \begin{minipage}[b]{0.45\textwidth}
	\begin{figure}[H] \begin{center}
	\includegraphics[width=\textwidth]{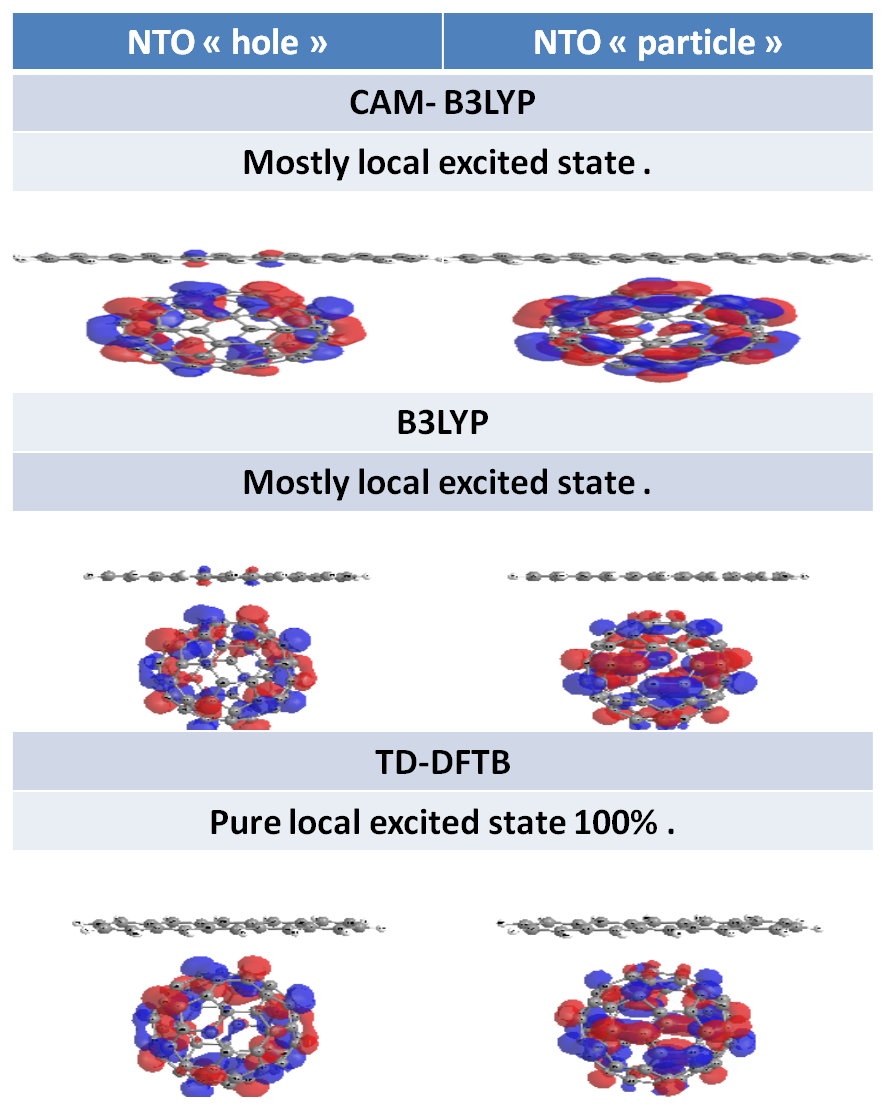}\\
	\scriptsize{c) Local excited state on  $\text{C}_{\text{60}}$ }
	\end{center} \end{figure}
    \end{minipage}
    \caption{The natural transition orbitals for excited states of model-2 obtained by different functionals and DFTB.}\label{fig:6}
    \end{center} \end{figure}
\end{widetext}

Figure \ref{fig:6} shows us the natural transition orbitals of the charge transfer state, the local excited state on pentacene, and the local excited state on fullerene of model-1 obtained with different functionals. It is worth mentioning here that when we go from a functional without range-separated hybrid to the functional with a range-separated hybrid, delocalization of the local excited state increases instead of decreasing. Probably that due to the strong intermolecular $\pi$ to $\pi$ interactions, there happens a critical hybridization between the pentacene and $\text{C}_{\text{60}}$ orbitals. As an outcome, the lowest CT state is considerably high in energy and blends with local states \cite{ZSYACB14}.

\onecolumngrid
\twocolumngrid

\subsection{Model-2, the central ring of pentacene centered over a 6-membered ring of $\text{C}_{\text{60}}$}
\begin{figure}[H]
\begin{center}
\includegraphics[width=\linewidth]{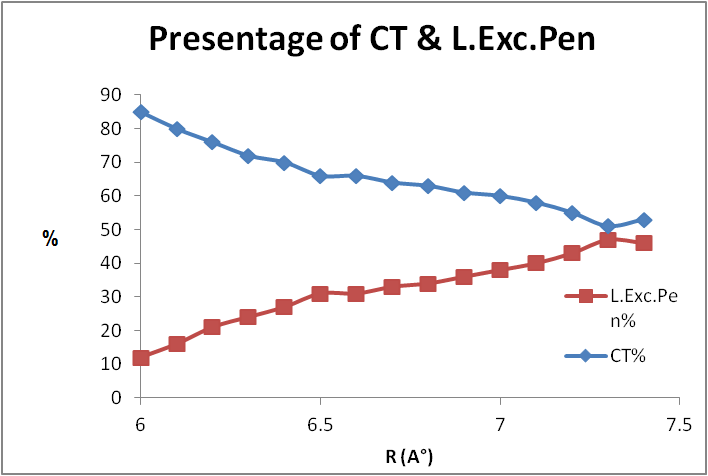}
\caption{\label{fig:7}The relationship between the distance and the percentage of each state at the same excitation by using CAM-B3LYP functional in model-2.} 
\end{center}
\end{figure}

The calculations of the excited states of model-2 by using range-separated hybrid with CAM-B3LYP functional at  6-31G(d,p) basis set level with empirical dispersion revealed  that the lowest excited state at 2.126 eV is a mixed state between charge transfer state from pentacene to $\text{C}_{\text{60}}$ (HOMO to LUMO) with percentage 64\% and local excited state on pentacene from HOMO to LUMO+3 with percentage 33\%. This percentage of mixing will be changed when we change the distance between the donor and the acceptor. The percentage of charge transfer state will decrease when increasing the distance between the donor and the acceptor that is a clear evidence that the charge transfer state depends on the strength of $\pi$-orbital stacking forces between the two fragments. Fig. \ref{fig:7} show us the relationship between the distance and the percentage of each state at the same excitation. 

Whereas, the lowest excited state of the same model by using B3LYP functional at 1.192 eV is a pure charge transfer state from HOMO to LUMO. With the same functional for the same model, the pure local excited state on pentacene from HOMO to LUMO+3 will be found at 1.919 eV and the excited state at 2.065 eV is mostly local excited state on fullerene which include transitions from HOMO-1, HOMO-1, HOMO-3, and HOMO-2 to LUMO, LUMO+1, LUMO+2, and LUMO+1 respectively whereas, the pure local excite state on fullerene by using CAM-B3LYP functional can be found at 2.539 eV and include six transitions from (HOMO - HOMO-5) to (LUMO - LUMO+2).
The image is quite different in DFTB calculations, the charge transfer state, the local excited state on pentacene, and the local excited state on fullerene all of them are a pure states. All the states for model-2 calculated by different functionals are shown in the table \ref{tab:4}.

\begin{figure}[H]
\begin{center}
\includegraphics[width=\linewidth]{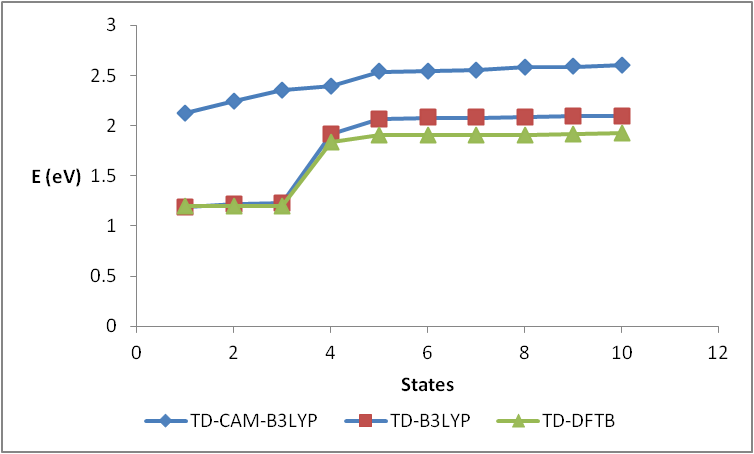}
\caption{\label{fig:8} The difference in energy between the functionals  for calculating the excited states in model-2.} 
\end{center}
\end{figure}

\begin{table}[htbp]
  \centering
  \caption{The local excitation state on pentacene (L. Exc. Pen.), the local excitation state on $\text{C}_{\text{60}}$ (L. Exc. $\text{C}_{\text{60}}$. S), and the charge transfer state from pentacene to $\text{C}_{\text{60}}$ in model-2 by using different methods.  All energies are given in eV.}
    \begin{tabular}{cccc}
    \toprule
    \multicolumn{4}{c}{\textbf{MODEL\_2}} \\
    \textbf{method} & \textbf{L. Exc. Pen.} & \textbf{L. Exc. C60} & \textbf{C.T.} \\
    \hline
    \text{OPT-ωB97XD \footnotemark[1]} & \text{2.090} & \text{2.520} & \text{2.090} \\
    \text{CAM-B3LYP\_D} & \text{2.397} & \text{2.539} & \text{2.126} \\
    \text{B3LYP} & \text{1.919} & \text{2.065} & \text{1.192} \\
    \text{TDDFTB} & \text{1.834} & \text{1.902} & \text{1.194} \\
    \toprule
    \end{tabular}%
    \footnotetext[1]{Taken from Ref.~\cite{ZSYACB14}.}
  \label{tab:4}%
\end{table}%
Table \ref{tab:4} shows us in clear manner there is no significant difference between optimized tuned range-separated hybrid functional (OPT-$\omega$B97XD) and CAM-B3LYP in the calculation of the charge transfer state, local excited state on pentacene, and local excited state on fullerene but the image is quite different when we go from functional with range separated hybrid to functional without range separated hybrid for example, the difference reach 1 eV in the charge transfer state energy between OPT-$\omega$B97XD and B3LYP functional but the difference between the functional with RSH and the functional without RSH in the local excited state on pentacene and local excited state on fullerene are around 0.5 eV and 0.2 respectively. The difference in energy between the functionals  for calculating the excited states in model-2 have been shown in Fig \ref{fig:8}.

At this point, we can say that the main difference between model-1 and model-2 is that the lowest excited state in model-1 is mostly a local excited state on pentacene from HOMO to LUMO+3 with percentage 74\% and small percentage 22\% of charge transfer state from pentacene to $\text{C}_{\text{60}}$ ( HOMO to LUMO) whereas the image in the model-2 is quite opposite, it is mostly charge transfer state from pentacene to $\text{C}_{\text{60}}$ ( HOMO to LUMO) with percentage 64\% and small percentage 33\% of local excited state on pentacene from HOMO to LUMO+3. We can attributed this change in the behavior in the model-2 to the strong   $\pi$-orbital stacking between the hexagonal rings in both pentacene and fullerene due to overlaps.

\onecolumngrid
\begin{widetext}
\begin{figure}[H]
\begin{center}
\includegraphics[width=\linewidth]{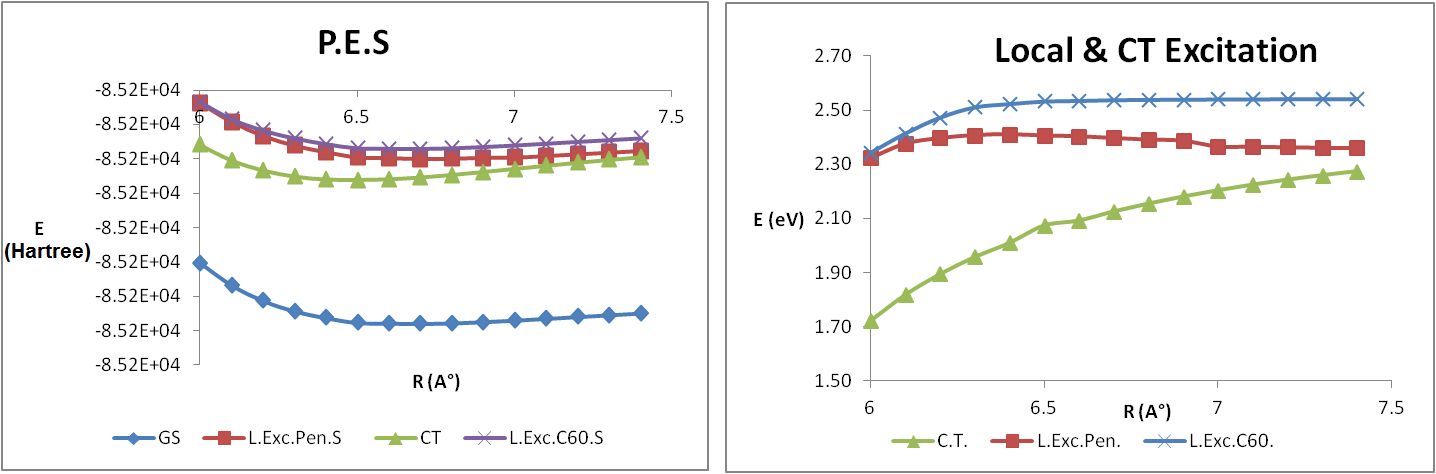}
\caption{\label{fig:9}Potential energy surfaces of the ground state (GS), the local excitation state on pentacene (L. Exc. Pen. S), the local excitation state on $\text{C}_{\text{60}}$ (L. Exc. $\text{C}_{\text{60}}$. S), and the charge transfer state from pentacene to $\text{C}_{\text{60}}$ in model-2 by using CAM-B3LYP with empirical dispersion GD3.} 
\end{center}
\end{figure}
\end{widetext}
\twocolumngrid
The potential energy surface of the ground state, the local excited state on pentacene, the local excited state on fullerene, and the charge transfer state have been also studied for model-2 by using CAM-B3LYP functional. These surfaces have been illustrated in Fig. \ref{fig:9}.  It is worth mentioning here, that there is an avoided crossing region at around 6.0 \AA \ between the potential energy surface of local excited state on pentacene and local excited state on fullerene. Also, there is a funnel region at around 7.5 \AA \ between the potential energy surface of the local excited state on pentacene and the potential energy surface of charge transfer state.

\onecolumngrid
\begin{widetext}
\begin{figure}[H] \begin{center}
    \begin{minipage}[b]{0.45\textwidth}
	\begin{figure}[H] \begin{center}
	\includegraphics[width=\textwidth]{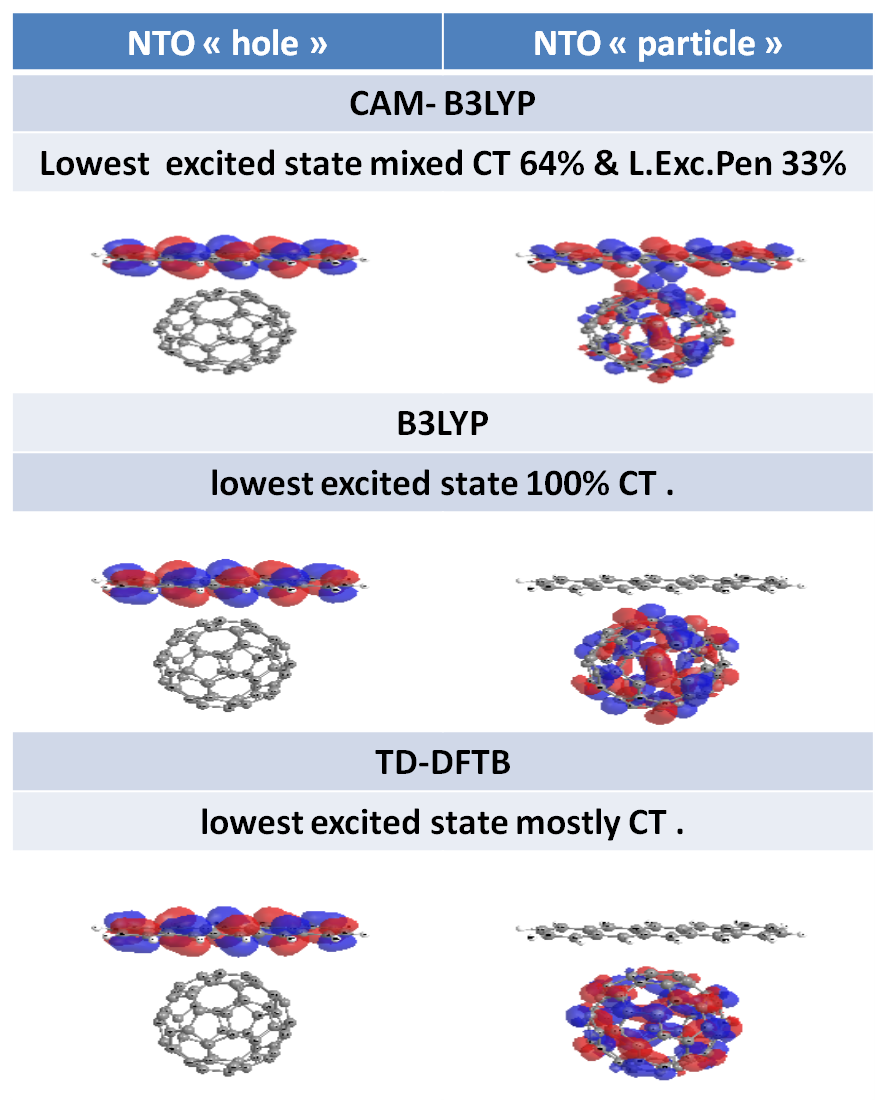}\\
	\scriptsize{a) Charge transfer state}
	\end{center} \end{figure}
    \end{minipage}
    \begin{minipage}[b]{0.45\textwidth}
	\begin{figure}[H] \begin{center}
	\includegraphics[width=\textwidth]{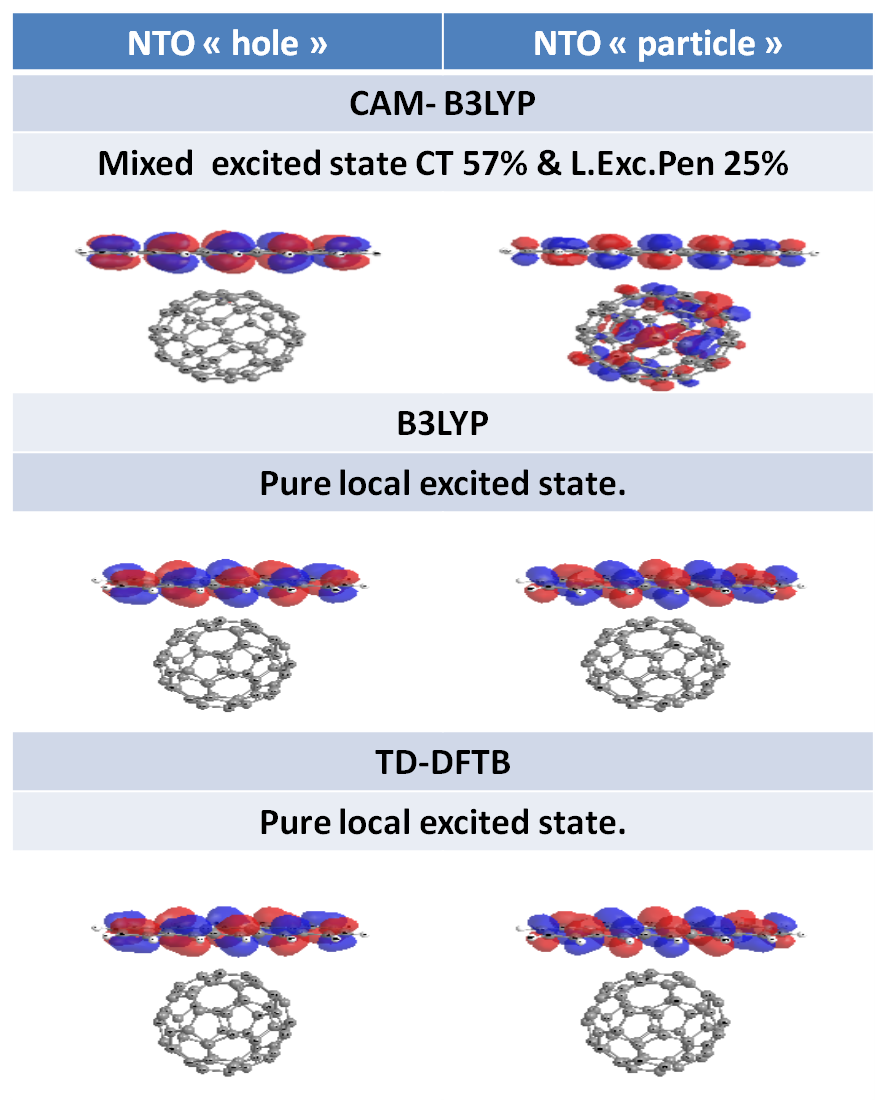}\\
	\scriptsize{b) Local excited state on pentacene}
	\end{center} \end{figure}
     \end{minipage}
     \begin{minipage}[b]{0.45\textwidth}
	\begin{figure}[H] \begin{center}
	\includegraphics[width=\textwidth]{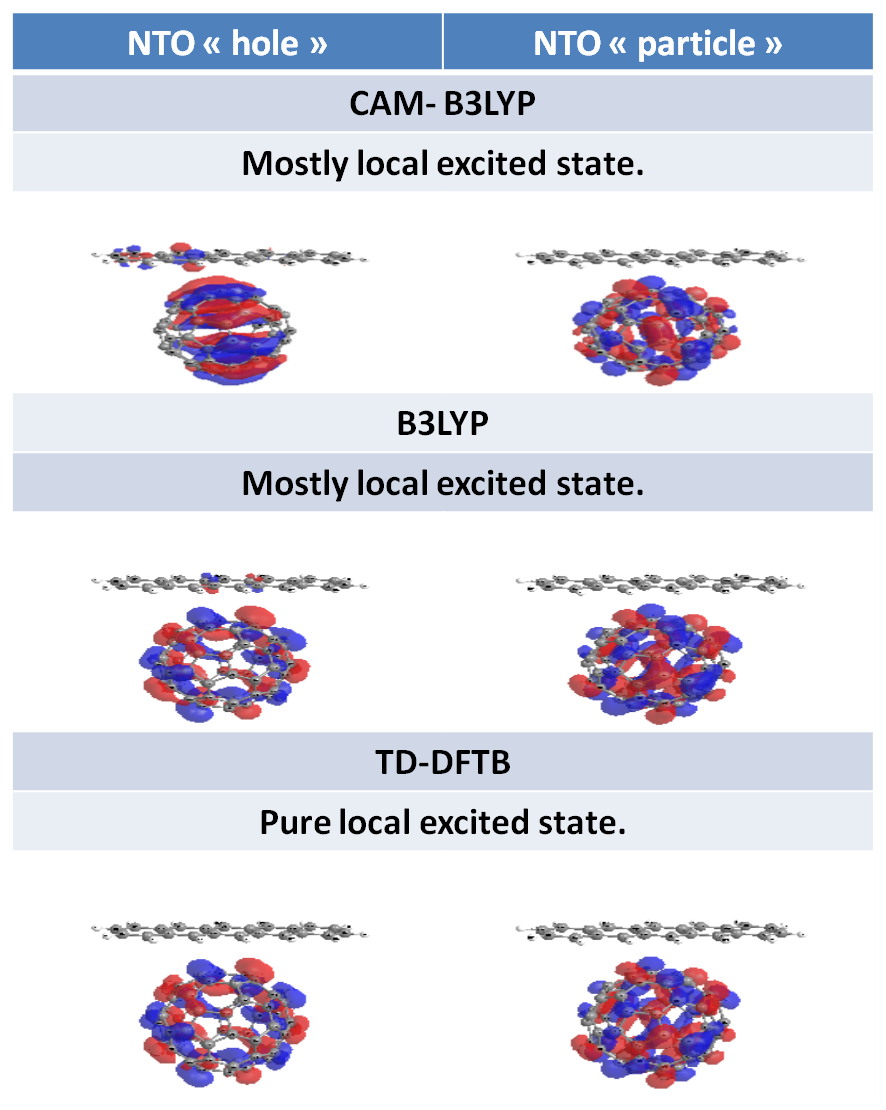}\\
	\scriptsize{c) Local excited state on  $\text{C}_{\text{60}}$ }
	\end{center} \end{figure}
    \end{minipage}
    \caption{The natural transition orbitals for excited states of model-2 obtained by different functionals.}\label{fig:10}
    \end{center} \end{figure}
\end{widetext}
\twocolumngrid
Figure \ref{fig:10} shows us the natural transition orbitals of the charge transfer state, the local excited state on pentacene, and the local excited state on fullerene of model-2 obtained by different functionals. It is worth to mentioned here when we go from functional without range-separated hybrid to the functional with a range-separated hybrid, a significant drop in the percentage of charge transfer happened from 100\% to 64\% and delocalization of the local excited state happened. Probably that due to the strong intermolecular $\pi$ to $\pi$ interactions, there happens a critical hybridization between the pentacene and $\text{C}_{\text{60}}$ orbitals. As an outcome, the lowest CT state is considerably brought up in energy and blends with local states \cite{ZSYACB14}. This drop in the percentage of charge transfer state in model-1 (68\%) is more than model-2 (36\%) that often due to the model-1 is weaker than model-2 in $\pi$-orbital stacking between the hexagonal rings in both pentacene and fullerene compare with $\pi$-orbital stacking between hexagonal ring in pentacene and pentagonal ring in fullerene.
Finally to complete the image of the model-2, the molecular orbitals composition of the frontier orbitals by each functional have been illustrated in the Table \ref{tab:5}. 

\onecolumngrid
\begin{widetext}
\begin{table}[H]
  \centering
  \caption{The molecular orbitals composition of each fragment in the model-2 by using different functionals}
    \begin{tabular}{ccccccccccccr}
    \toprule
    \multicolumn{13}{c}{\textbf{MODEL\_2}} \\
     \toprule
    \textbf{} & \multicolumn{12}{c}{\textbf{CAM-B3LYP}} \\
    orbitals & H-5   & H-4   & H-3   & H-2   & H-1   & H     & L     & L+1   & L+2   & L+3   & L+4   & \multicolumn{1}{c}{L+5} \\
    pentacene & 0     & 1     & 1     & 2     & 3     & 98    & 2     & 0     & 1     & 93    & 0     & \multicolumn{1}{c}{1} \\
    fullerene & 100   & 99    & 99    & 98    & 97    & 2     & 98    & 100   & 99    & 7     & 100   & \multicolumn{1}{c}{99} \\
    \textbf{} & \multicolumn{12}{c}{\textbf{B3LYP}} \\
    pentacene & 0     & 1     & 3     & 4     & 3     & 98    & 2     & 0     & 1     & 93    & 0     & \multicolumn{1}{c}{1} \\
    fullerene & 100   & 99    & 97    & 96    & 97    & 2     & 98    & 100   & 99    & 7     & 100   & \multicolumn{1}{c}{99} \\
    \textbf{} & \multicolumn{12}{c}{\textbf{DFTB}} \\
    pentacene & 0     &       & 0     & 0     & 0     & 100   & 0     & 1     & 0     & 99    & 0     & \multicolumn{1}{c}{0} \\
    fullerene & 100   &       & 100   & 100   & 100   & 0     & 100   & 99    & 100   & 1     & 100   & \multicolumn{1}{c}{100} \\
    \toprule
    \end{tabular}%
  \label{tab:5}%
\end{table}%
\end{widetext}
\twocolumngrid

\subsection{Model-R, the central 6,13 carbons in pentacene centred over  a the $\pi$ bond between two fused 6-membered rings in $\text{C}_{\text{60}}$}

The calculations of the excited states of model-R by using range-separated hybrid functional (CAM-B3LYP)  at 6-31G(d,p) basis set level with empirical dispersion revealed that the image is different from model-1 and model-2 because the lowest excited state at 2.247 eV  in model-R is a pure local excited state on pentacene from HOMO to LUMO+3. By the same functional, the second excited state is a pure charge transfer state from HOMO to LUMO+1 with percentage 95\%  and from HOMO to LUMO+5 with percentage 5\%. The local excited state on fullerene can be seen clearly in the fifth excited state like model-1 and model-2  which include transitions from HOMO-2, HOMO-1, HOMO-3, HOMO-6, and HOMO-5 to LUMO +2, LUMO+1, LUMO, LUMO +1, and LUMO +1 with precentage 46\%, 35\%, 9\%, 2\%, and 2\% respectively.
Whereas, the calculations by using B3LYP functional showed the lowest excited state at 1.347 eV is a pure charge transfer state from HOMO to LUMO+1. With the same functional for the same model, the pure local excited state on pentacene from HOMO to LUMO+3 will be found at 1.999 eV and the excited state at 2.133 eV is also a pure local excited state on fullerene which include transitions from HOMO-1 to LUMO with precentage 88\% and from HOMO-5 to LUMO+2 to LUMO with precentage 8\%. 
The image is not very different in DFTB calculations, the charge transfer state and the local excited state on pentacene each one of them is a pure state  but the local excited state on fullerene is a linear combination several local states. All the states for model-R calculated by different functionals are shown in the Table \ref{tab:6} .

\begin{table}[htbp]
  \centering
  \caption{The local excitation state on pentacene (L. Exc. Pen.), the local excitation state on $\text{C}_{\text{60}}$ (L. Exc. $\text{C}_{\text{60}}$. S), and the charge transfer state from pentacene to $\text{C}_{\text{60}}$ in model-R by using different methods.}
    \begin{tabular}{rccc}
    \toprule
          & \textbf{MODEL\_R} & \textbf{} & \textbf{} \\
   
    \multicolumn{1}{c}{\textbf{method}} & \textbf{L. Exc. Pen.} & \textbf{L. Exc. $\text{C}_{\text{60}}$} & \textbf{C.T.} \\
    \hline
    \multicolumn{1}{c}{\text{CAM-B3LYP\_D}} & \text{2.373} & \text{2.602} & \text{2.381} \\
    \multicolumn{1}{c}{\text{B3LYP}}        & \text{1.999} & \text{2.133} & \text{1.347} \\
    \multicolumn{1}{c}{\text{TDDFTB}}       & \text{1.897} & \text{1.970} & \text{1.295} \\
    \toprule
    \end{tabular}%
  \label{tab:6}%
\end{table}%

\begin{figure}[H]
\begin{center}
\includegraphics[width=\linewidth]{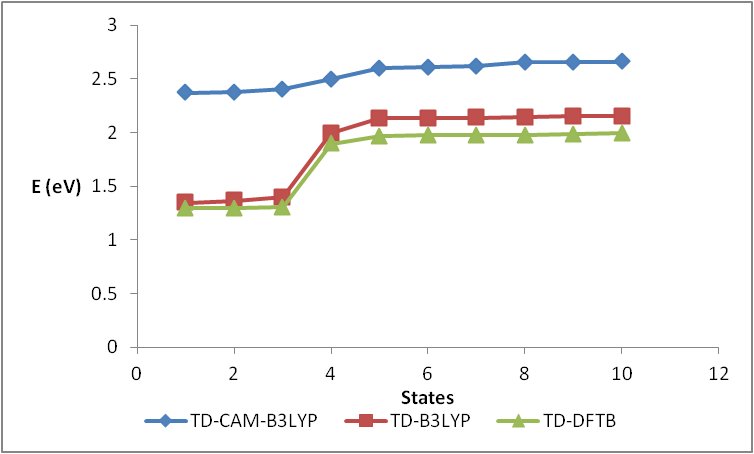}
\caption{\label{fig:11} The difference in energy between the functionals  for calculating the excited states in model-R.} 
\end{center}
\end{figure}

Table \ref{tab:6} shows us in clear manner there is no significant difference between the behavior of B3LYP and DFTB in the calculation of the charge transfer state, local excited state on pentacene, and local excited state on fullerene but the image is quite different when we go from functional with range separated hybrid to functional without range separated hybrid for example, the difference reach 1 eV in the charge transfer state energy between CAM-B3LYP and B3LYP functional but the difference between the functional with RSH and the functional without RSH in the local excited state on pentacene and local excited state on fullerene are around 0.5 eV. The difference in energy between the functionals for calculating the excited states in model-R have been shown in Fig. \ref{fig:11}.

At this point, we can say that the main difference between model-R and both model-1 and model-2 is that the lowest excited state by using CAM-B3LYP functional in both model-1 and model-2 is a mixed state between the charge transfer state and the local excited state on pentacene whereas the image in the model-R is quite opposite, it is a pure local excited state on pentacene from HOMO to LUMO+3.
Here, I want to mention that when we go from mode-2 which have strongest $\pi$-orbital stacking between the hexagonal rings in both pentacene and fullerene due to the symmetry passing through the model-1 which is $\pi$-orbital stacking between the hexagonal rings in pentacene and pentagonal ring in fullerene less than model-2 down to the model-R which is the poorest $\pi$-orbital stacking, the percentage of local excited state on pentacene in the lowest excited state increase from 33\% in model-2 to 64\% in model-1 to 100 \% in the model-R. And the percentage of the charge transfer state in the lowest excited state take the opposite order it increase from 0\% in the model-R to the 24\% in the model-1 to the 64\% in the model-2 due to the increasing of $\pi$-orbital stacking.  

\begin{widetext}
\begin{figure}[H]
\begin{center}
\includegraphics[width=\linewidth]{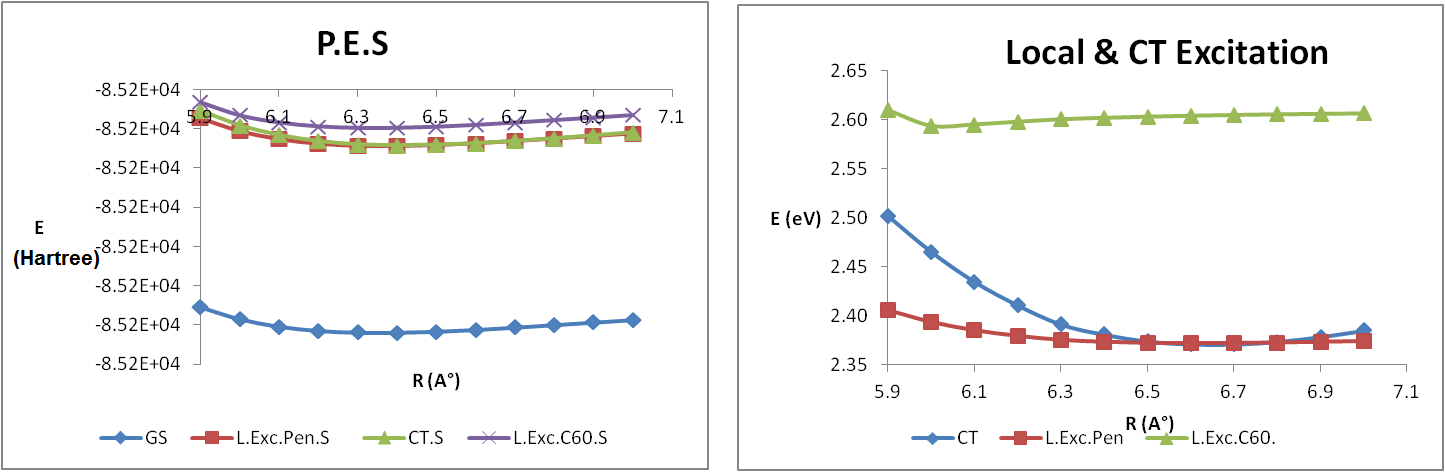}
 \caption{\label{fig:12}The potential energy surfaces of the ground state (GS), the local excitation state on pentacene (L. Exc. Pen. S), the local excitation state on $\text{C}_{\text{60}}$ (L. Exc. $\text{C}_{\text{60}}$. S), and the charge transfer state from pentacene to $\text{C}_{\text{60}}$ in model-R by using CAM-B3LYP with empirical dispersion GD3.} 
\end{center}
\end{figure}
\end{widetext}

The potential energy surfaces in eV of the ground state (GS), the local excitation state on pentacene, the local excited state on $\text{C}_{\text{60}}$, and the charge transfer state from pentacene to $\text{C}_{\text{60}}$ in model-R by using CAM-B3LYP with empirical dispersion GD3 have been shown in Fig \ref{fig:12}.

\begin{widetext}
\begin{figure}[H] \begin{center}
    \begin{minipage}[b]{0.45\textwidth}
	\begin{figure}[H] \begin{center}
	\includegraphics[width=\textwidth]{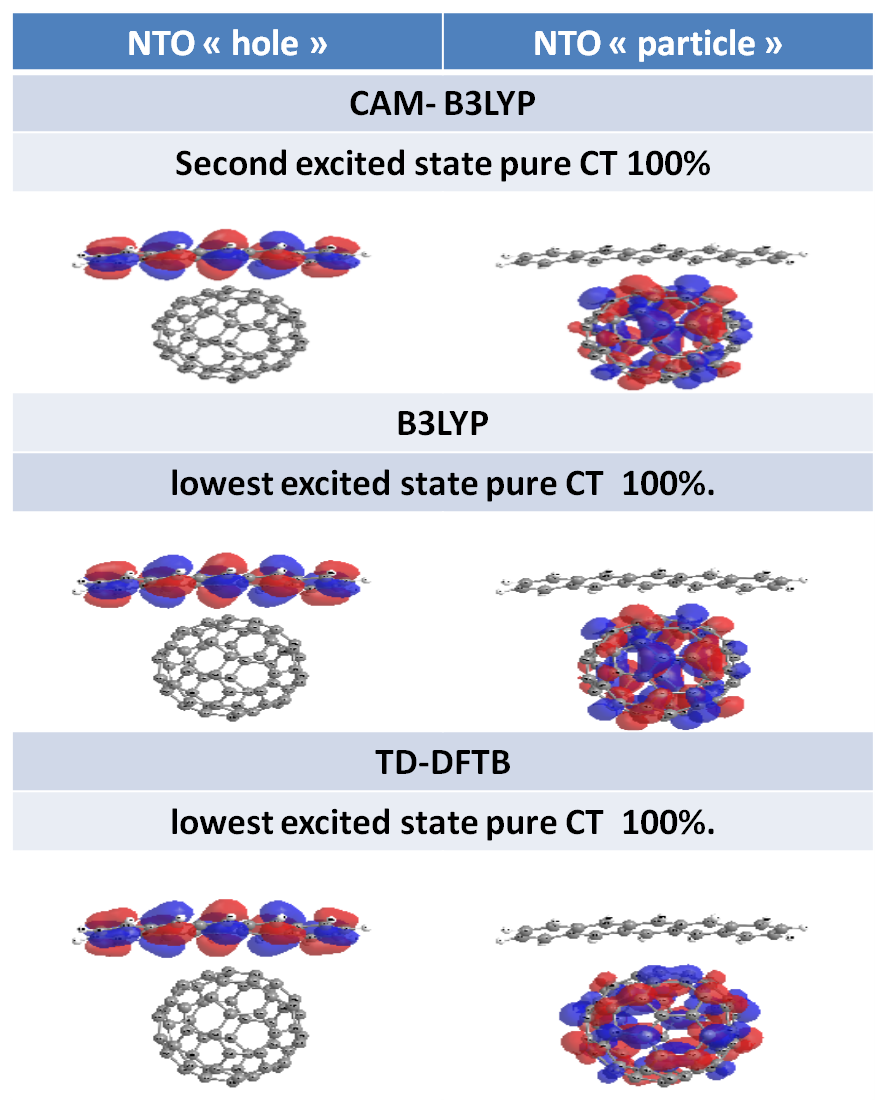}\\
	\scriptsize{a) Charge transfer state}
	\end{center} \end{figure}
    \end{minipage}
    \begin{minipage}[b]{0.45\textwidth}
	\begin{figure}[H] \begin{center}
	\includegraphics[width=\textwidth]{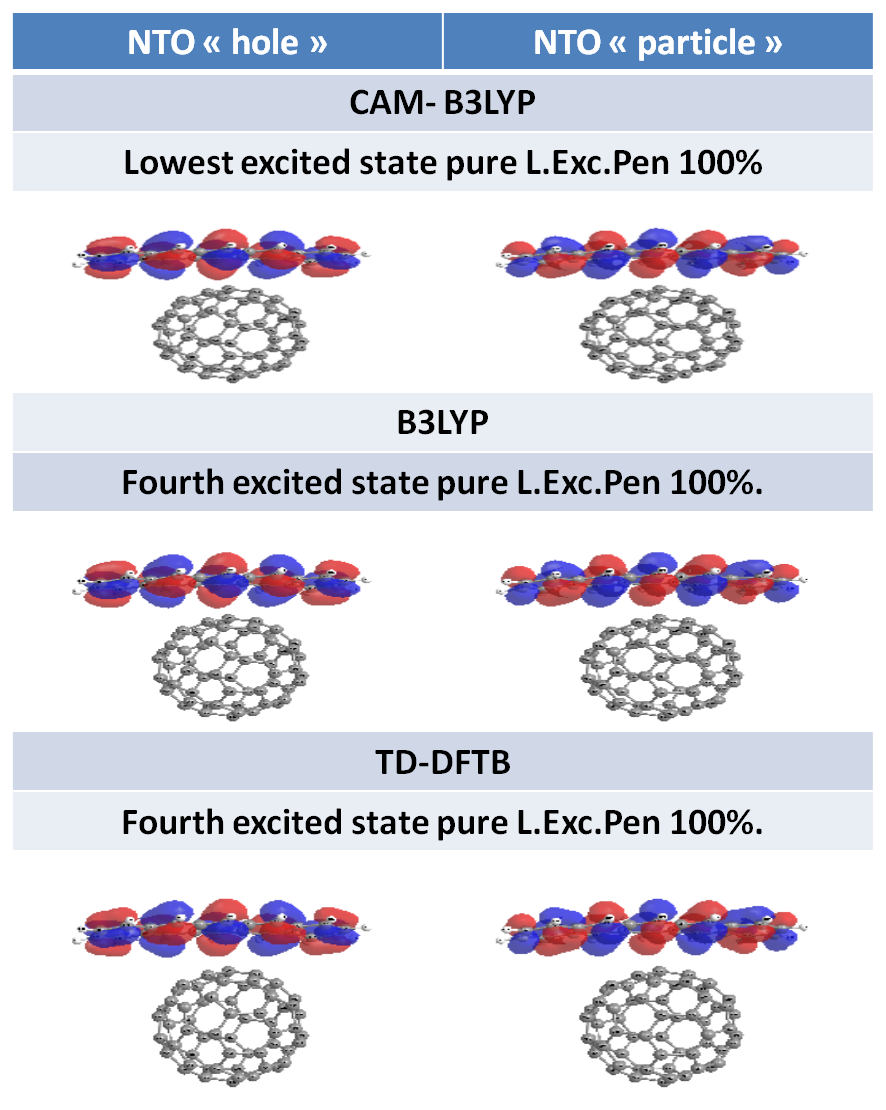}\\
	\scriptsize{b) Local excited state on pentacene}
	\end{center} \end{figure}
     \end{minipage}
     \begin{minipage}[b]{0.45\textwidth}
	\begin{figure}[H] \begin{center}
	\includegraphics[width=\textwidth]{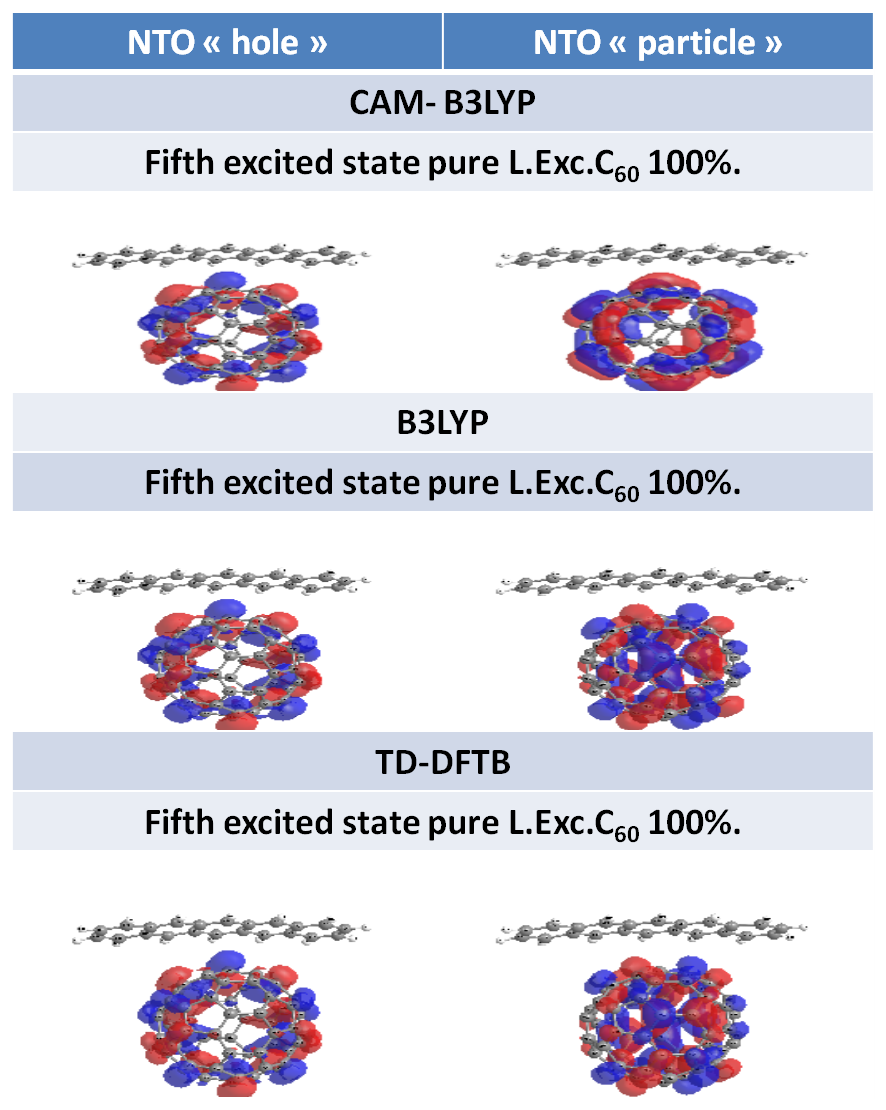}\\
	\scriptsize{c) Local excited state on  $\text{C}_{\text{60}}$ }
	\end{center} \end{figure}
    \end{minipage}
    \caption{The natural transition orbitals for excited states of model-R obtained by different functionals.}\label{fig:13}
    \end{center} \end{figure}
\end{widetext}

\onecolumngrid
\twocolumngrid 
Figure \ref{fig:13} shows us the natural transition orbitals of the charge transfer state, the local excited state on pentacene, and the local excited state on fullerene of model-R obtained by different functionals. It is worth mentioning here that in the model-R, the local excited state on pentacene and the local excited state on fullerene with different functionals and DFTB are purely local excited states also the charge transfer state with different functionals and DFTB in the model-R is a purely charge transfer state. (see in the Fig. \ref{fig:13}) .

\onecolumngrid
\twocolumngrid
\subsection{Model-P, the central 6,13 carbons in pentacene centred over  a the $\pi$ bond between two fused 6-membered rings in $\text{C}_{\text{60}}$}

The calculations of the excited states of model-P by using different functionals revealed that the image is different from all previous models because we can not find any local excitation on pentacene within the first 30 excited states. Perhaps this happens due to the frontier molecular orbital compositions are almost those of fullerene as shown in the Table \ref{tab:7}.

\onecolumngrid
\begin{widetext}
\begin{table}[htbp]
  \centering
  \caption{The frontier molecular orbital compositions of the model-P by using different methods}
    \begin{tabular}{ccccccccccccr}
    \toprule
    \multicolumn{13}{c}{\textbf{MODEL\_P}} \\
    \toprule
    \textbf{} & \multicolumn{12}{c}{\textbf{CAM-B3LYP}} \\
    orbitals & H-5   & H-4   & H-3   & H-2   & H-1   & H     & L     & L+1   & L+2   & L+3   & L+4   & \multicolumn{1}{c}{L+5} \\
    pentacene & 90    & 73    & 6     & 0     & 11    & 2     & 1     & 1     & 3     & 2     & 3     & \multicolumn{1}{c}{26} \\
    fullerene & 10    & 27    & 94    & 100   & 89    & 98    & 99    & 99    & 97    & 98    & 97    & \multicolumn{1}{c}{74} \\
    \textbf{} & \multicolumn{12}{c}{\textbf{B3LYP}} \\
    pentacene & 63    & 90    & 5     & 0     & 22    & 2     & 1     & 1     & 3     & 1     & 25    & \multicolumn{1}{c}{2} \\
    fullerene & 37    & 10    & 95    & 100   & 78    & 98    & 99    & 99    & 97    & 99    & 75    & \multicolumn{1}{c}{98} \\
    \textbf{} & \multicolumn{12}{c}{\textbf{DFTB}} \\
    pentacene & 85    & 3     & 1     & 0     & 0     & 4     & 0     & 0     & 1     & 0     & 0     & \multicolumn{1}{c}{0} \\
    fullerene & 15    & 97    & 99    & 100   & 100   & 96    & 100   & 100   & 99    & 100   & 100   & \multicolumn{1}{c}{100} \\
     \toprule
    \end{tabular}%
  \label{tab:7}%
\end{table}%
\end{widetext}
\twocolumngrid

The calculations of the excited states obtained by using range-separated hybrid functional (CAM-B3LYP) and 6-31G(d,p) basis set with empirical dispersion showed 
the lowest excited state at 2.533 eV is a mostly local excited state on fullerene from HOMO to LUMO with percentage 91\% and from HOMO-1 to LUMO+1 with percentage 3\%. By the same functional, the second excited state is a mixed state between the charge transfer state from HOMO-3 to LUMO+1 with percentage 14\%  and local excited state on fullerene from HOMO to LUMO+1 with percentage 79\%.
Whereas, the calculations by using B3LYP functional showed the same behavior, the lowest excited at 2.018 eV is mostly local excited state on fullerene from HOMO to LUMO+1 with percentage 78\%. With the same functional for the same model, the second excited state is a mixed state between the charge transfer state from HOMO-1 to LUMO with percentage 19\%  and local excited state on fullerene from HOMO to LUMO+1 with percentage 79\%.

DFTB calculations fail to describe the charge transfer state in this model. We did not find any charge transfer state in the first 30 excited states, with the exception of a small portion (3\% percentage) in the second excited state. The lowest excited state according DFTB calculation is a local excited state on fullerene from HOMO to LUMO.  
The local excited state on fullerene, and the charge transfer state from pentacene to fullerene of model-P calculated by different functionals are shown in the table \ref{tab:8}.

\begin{table}[htbp]
  \centering
  \caption{The local excitation state on $\text{C}_{\text{60}}$ (L. Exc. $\text{C}_{\text{60}}$. S), and the charge transfer state from pentacene to $\text{C}_{\text{60}}$ in model-P by using different methods. All energies are given in eV.}
    \begin{tabular}{ccc}
    \toprule
    \textbf{method} & \textbf{L. Exc. C60.} & \textbf{C.T.} \\
    \toprule
    \text{CAM-B3LYP\_D} & \text{2.533} & \text{2.611} \\
    \text{B3LYP} & \text{2.018} & \text{2.115} \\
    \text{TDDFTB} & \text{1.857} & \text{1.939} \\
    \toprule
    \end{tabular}%
  \label{tab:8}%
\end{table}%

\begin{figure}[H]
\begin{center}
\includegraphics[width=\linewidth]{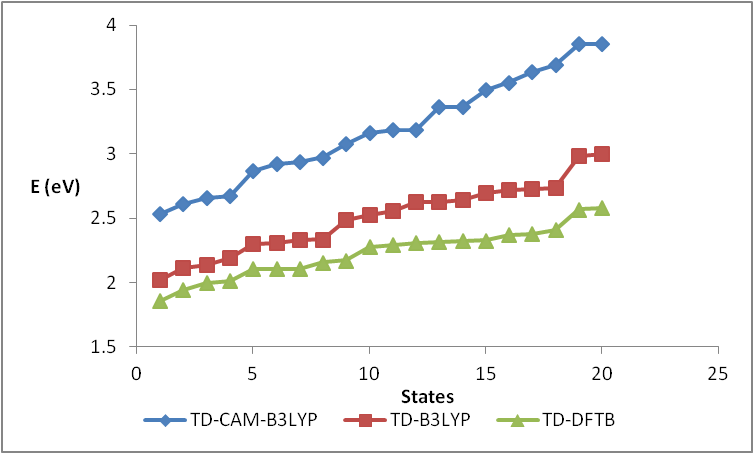}
\caption{\label{fig:15} The difference in energy between the functionals  for calculating the excited states in model-P.} 
\end{center}
\end{figure}

Table \ref{tab:8} shows that the difference between B3LYP and DFTB in the calculation of the local excited state on fullerene less than 0.2 eV but the different  will be increase when we go from functional with range-separated hybrid to functional without range-separated hybrid for example, the difference around  0.5 eV in the charge transfer state energy and the local excited state energy on fullerene between CAM-B3LYP and B3LYP functional. The difference in energy between the functionals for calculating the excited states in model-R have been shown in Fig. \ref{fig:15}.

In the second excited state we studied the relation between the percentage of charge transfer state and the distance between the donor and the acceptor, We found that the percentage of charge transfer state will increase when increasing the distance between the donor and the acceptor until reach 5.4 \AA \ after that the percentage will decrease whereas the behavior of the local excited state on fullerene in the second excited state will be opposite, the precentage will decrease until reach 5.4 \AA \ then it will increase that is a clear evidence that the charge transfer state depends on the $\pi$-orbital stacking between the two fragments.

\begin{widetext}
\begin{figure}[H] \begin{center}
    \begin{minipage}[b]{0.45\textwidth}
	\begin{figure}[H] \begin{center}
	\includegraphics[width=\textwidth]{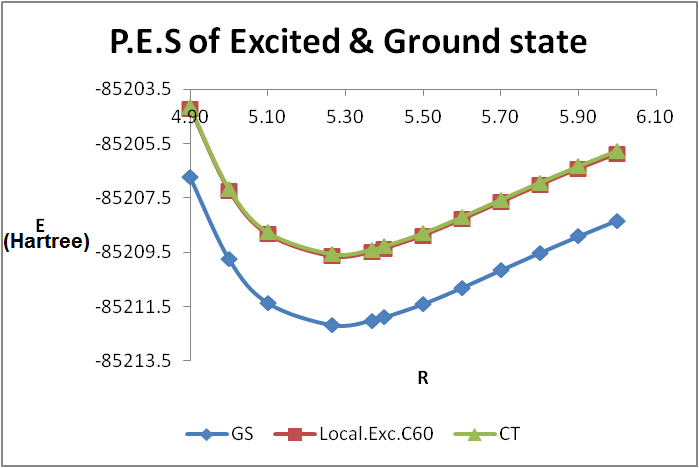}\\
	\end{center} \end{figure}
    \end{minipage}
    \begin{minipage}[b]{0.45\textwidth}
	\begin{figure}[H] \begin{center}
	\includegraphics[width=\textwidth]{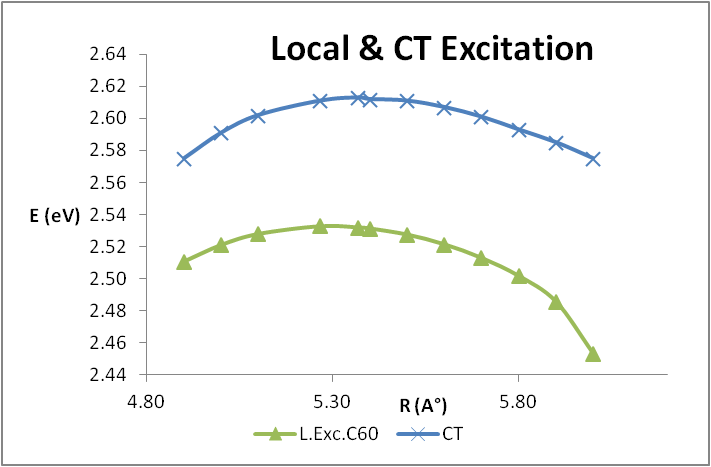}\\
	\end{center} \end{figure}
    \end{minipage}
    \begin{minipage}[b]{0.45\textwidth}
	\begin{figure}[H] \begin{center}
		\scriptsize{a) The potential energy surfaces in eV of the ground state (GS), the local excitation state on $\text{C}_{\text{60}}$ (L. Exc. $\text{C}_{\text{60}}$), and the charge transfer state from in model-P by using CAM-B3LYP .}
	\end{center} \end{figure}
    \end{minipage}
    \begin{minipage}[b]{0.45\textwidth}
	\begin{figure}[H] \begin{center}
	\scriptsize{b) The Local excited state on $\text{C}_{\text{60}}$ and the charge transfer state in model-P }
	\end{center} \end{figure}
    \end{minipage}
    \begin{minipage}[b]{0.45\textwidth}
	\begin{figure}[H] \begin{center}
	\includegraphics[width=\textwidth]{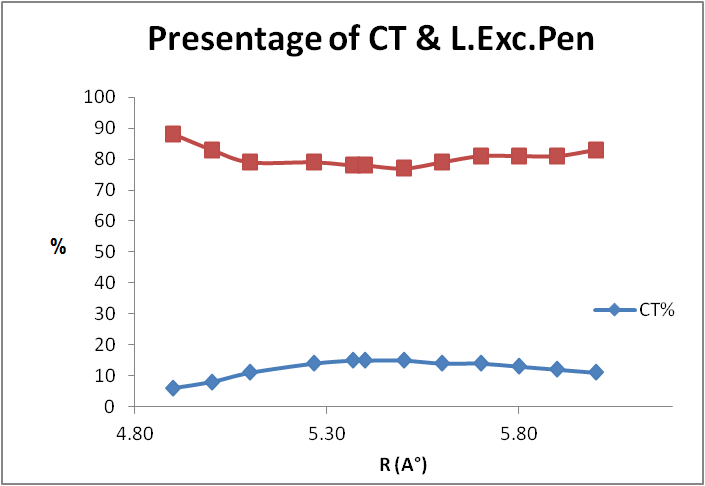}\\
	\scriptsize{c) The relationship between the distance and the percentage of each state at the same excitation by using CAM-B3LYP functional in model-P.}
	\end{center} \end{figure}
    \end{minipage}
    \caption{The potential energy curves in eV}\label{fig:17}
    \end{center} \end{figure}
    \end{widetext}

Fig. \ref{fig:17} (c) show us the relationship between the distance and the percentage of each state at the second  excited state. Also the potential energy surfaces in eV of the ground state (GS) and the local excited state on $\text{C}_{\text{60}}$, and the charge transfer state from pentacene to $\text{C}_{\text{60}}$ in model-P by using CAM-B3LYP with empirical dispersion GD3 have been studied and shown in Figs. \ref{fig:17} (a) and (b).

\twocolumngrid 
\onecolumngrid
\twocolumngrid

\begin{widetext}
\begin{figure}[H] \begin{center}
    \begin{minipage}[b]{0.45\textwidth}
	\begin{figure}[H] \begin{center}
	\includegraphics[width=\textwidth]{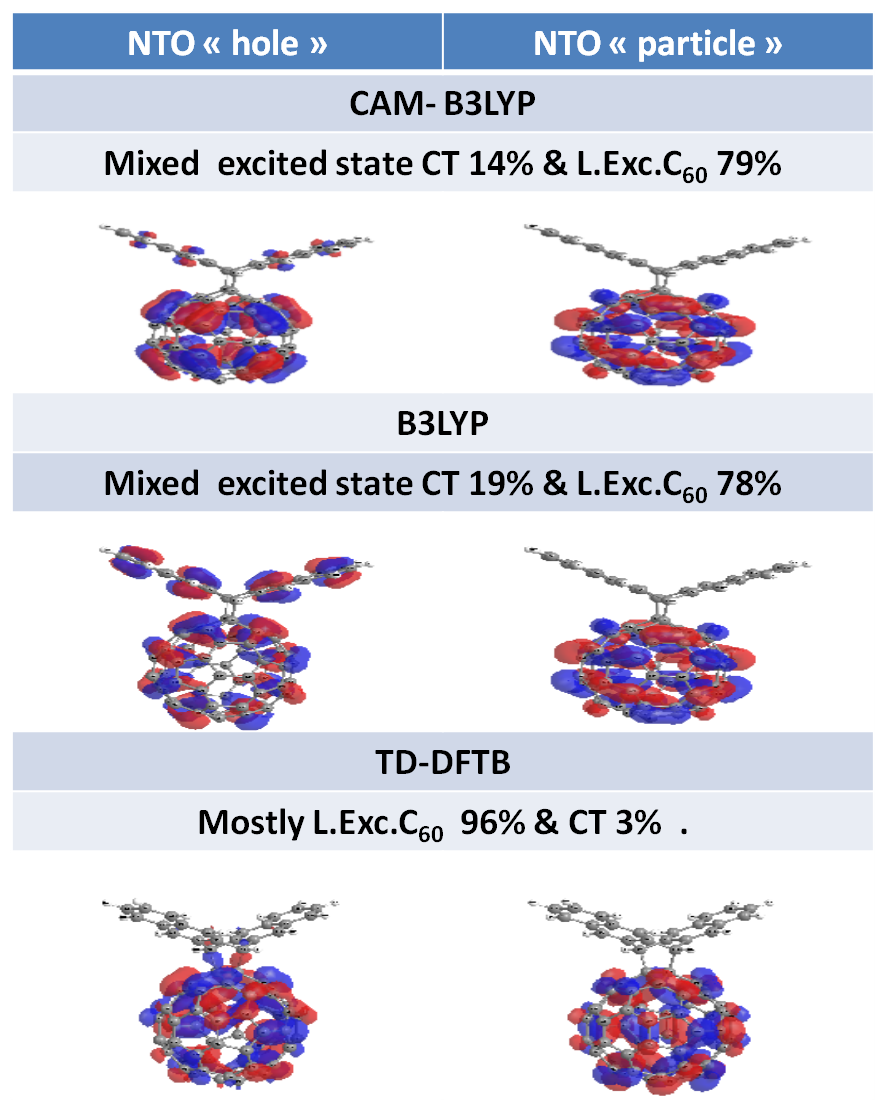}\\
	\scriptsize{a) Charge transfer state}
	\end{center} \end{figure}
    \end{minipage}
    \begin{minipage}[b]{0.45\textwidth}
	\begin{figure}[H] \begin{center}
	\includegraphics[width=\textwidth]{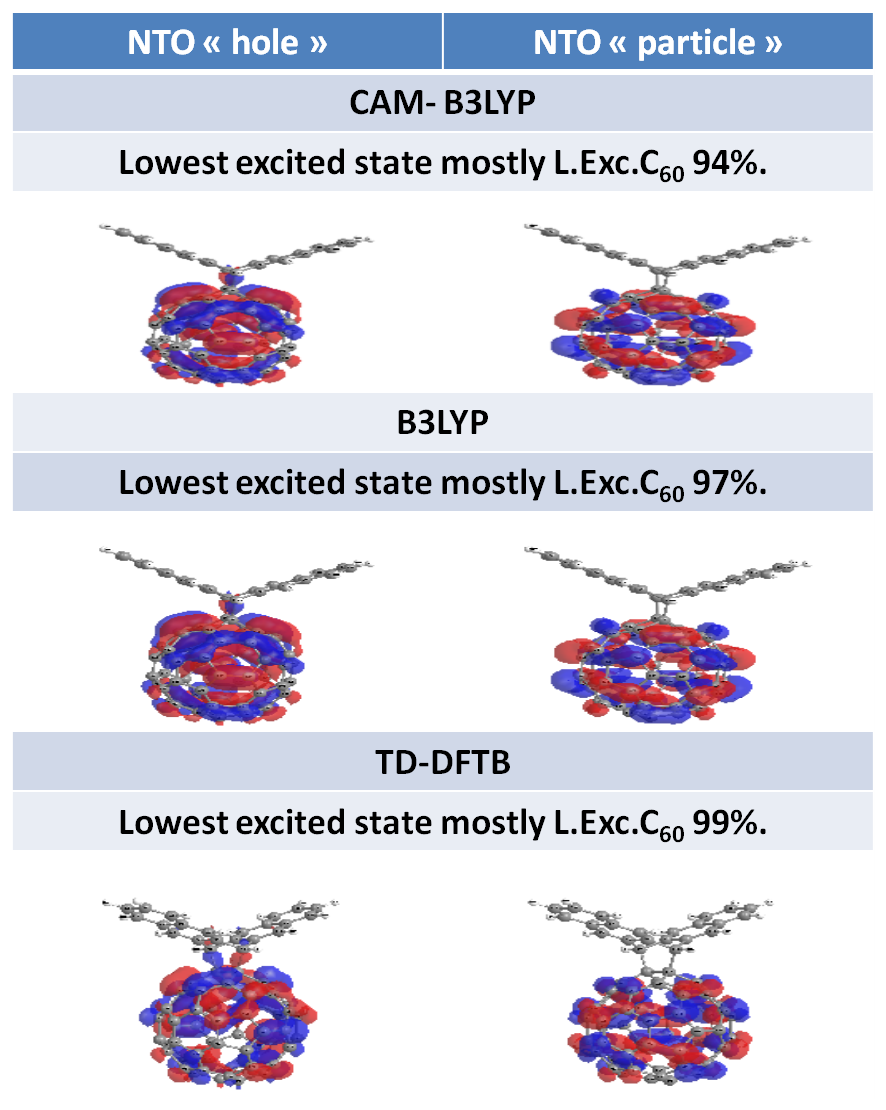}\\
	\scriptsize{c) Local excited state on  $\text{C}_{\text{60}}$ }
	\end{center} \end{figure}
    \end{minipage}
    \caption{The natural transition orbitals for excited states of model-P obtained by different functionals.}\label{fig:16}
    \end{center} \end{figure}
    \end{widetext}

At this point, we can notice the main difference between model-P and all previous models that the local excited state on pentacene disappeared in model-P and the percentage of the charge transfer state in model-P is the lowest among all models. Probably when we go from mode-2 which have strongest $\pi$-orbital stacking between the hexagonal rings in both pentacene and fullerene due to the symmetry passing through the model-1 which is $\pi$-orbital stacking between the hexagonal rings in pentacene and pentagonal ring in fullerene less than model-2 to the model-R which it is  poorer $\pi$-orbital stacking than model-1 down to the model-P which is the poorest in $\pi$-orbital stacking among all the models where, the model-P just has only 
one charge transfer state with percentage 14\% in second excited state among 30 excited states calculated by using CAM-B3LYP functional. Figure \ref{fig:16} shows us the natural transition orbitals of the charge transfer state and the local excited state on fullerene of model-P obtained by different functionals.
\onecolumngrid
    \twocolumngrid
\section{Conclusion}
The potential energy surface of the model systems by using CAM-B3LYP functional with counterpoise corrections revealed that the binding energies between two fragments in model-1, model-2, and model-R are too close -0.3041 eV in model-1, -0.3045 in model-2, and -0.3628 in model-R. Whereas, the binding energy between two fragments in model-P twenty-three times larger than the binding energy of both model-1 and model-2 and fourteen times larger than the binding energy of model-R that it is normal due to the nature of $\pi$ bond that bind between pentacene and $\text{C}_{\text{60}}$ in the model-P compare with van der Waals forces that bind between the molecules in the model-1, model-2, and model-R. Also, there is difference around 0.2 eV between our calculations and the calculations of binding energy by Zhang \textit{et al.} for the model-1 and the model-2. This difference can be attributed to the counterpoise corrections that missed by Zhang \textit{et al.} calculations.

The obvious thing in the calculations of the lowest excited state by using range-separated hybrid functional of all models that the percentage of charge transfer state increase from 0\% in both model-R and model-P passing through 22\% in model-1 to reach to 64\% in model-2 depending on the strength of $\pi$-orbital stacking between the two fragments in the model.
whereas, the opposite behavior can be noticed with the local excited state in the lowest excited state, the percentage of the local excited state increase from 33\% in model-2 passing through 74\% in model-1 to reach 100\% in both model-R and model-P according to the weakness of $\pi$-orbital stacking between the two fragments in the model.

The results shows us in clear manner there is no significant difference between the behavior of B3LYP and DFTB in the calculation of the charge transfer state, local excited state on pentacene, and local excited state on fullerene but the image is quite different when we go from functional with range separated hybrid to functional without range separated hybrid for example, the difference reach to 1 eV in the charge transfer state energy between CAM-B3LYP and both B3LYP functional and DFTB methods  in the model-1, model-2, and model-R and decrease to 0.5 eV in the model-P.
But the difference between the functional with RSH and the functional without RSH in the local excited state on pentacene and local excited state on fullerene are around 0.5 eV.

\section*{Acknowledgements}
I would like to express my deep gratitude to Professor Mark E.\ Casida, 
my research supervisor, for his patient guidance, enthusiastic encouragement 
and useful critiques of this research work.and l would like to thank the 
French and Irakian governments for support through a Campus France scholarship.
I would also like to thank Profs.\ Roland Mitri\v{c}. And I would also like to acknowledge ORGAnic solar cell VOLTage by numerical computation ANR-12-MONU-0014-02. In the end, I do not want to forget to thank Dr. Hemanadhan Myneni for all his efforts and for useful discussions.
\bibliographystyle{myaip}
\bibliography{refs}
\end{document}